\documentclass[longauth]{aa} 

\usepackage{graphicx}
\usepackage{natbib}

\usepackage[varg]{txfonts}
\begin{document}
\title{Outflow forces in intermediate mass star formation}

\author{ 
T.A. van Kempen  \inst{1,2} \and
M.R. Hogerheijde \inst{1} \and
E.F. van Dishoeck \inst{1,3} \and
L. E. Kristensen \inst{1,4} \and
A. Belloche \inst{5} \and
P.D. Klaassen \inst{1,6} \and
S. Leurini \inst{5} \and
I. San Jose-Garcia \inst{1} \and 
A. Aykutalp \inst{7,8} \and
Y. Choi \inst{7,9} \and
A. Endo \inst{10,11} \and
W. Frieswijk \inst{12} \and
D. Harsono \inst{1} \and
A. Karska\inst{13,3,1} \and
E. Koumpia \inst{7,9} \and
N. van der Marel \inst{1} \and
Z. Nagy \inst{7,9} \and
J.-P. P{\'e}rez-Beaupuits \inst{5} \and
C. Risacher \inst{5,9} \and
R.J. van Weeren \inst{1,4} \and
F. Wyrowski \inst{5} \and
U.A. Y{\i}ld{\i}z \inst{1,14} \and
R. G\"usten\inst{5} \and
W. Boland \inst{1,15} \and
A. Baryshev \inst{7,9}
}

\institute{
$^1$ Leiden Observatory, Leiden University, Niels Bohrweg 2, 2333 CA, Leiden, the Netherlands \\ \email{kempen@strw.leidenuniv.nl}\\
$^2$ Joint ALMA Offices, Av. Alonso de Cordova 3107, Vitacura, Santiago, Chile \\
$^3$ Max-Planck-Institut f\"ur Extraterrestische Physik, Giessenbachstrasse 2, 85478, Garching, Germany \\
$^4$ Harvard-Smithsonian Center for Astrophysics, 60 Garden Street, Cambridge, MA 02138, USA\\
$^5$ Max-Planck-Institut f\"ur Radioastronomie, Auf dem H\"ugel 69, 53121, Bonn, Germany\\
$^6$ UK Astronomy Technology Center, Royal Observatory Edinburgh, Blackford Hill, Edinburgh EH9 3HJ, UK\\
$^7$ Kapteyn Institute, University of Groningen, Landleven 12, 9747 AD, Groningen, the Netherlands \\
$^8$ Scuola Normale Superiore, Piazza dei Cavalieri 7, 56126, Pisa, Italy\\
$^9$ Netherlands Institute for Space Research, Low Energy Astrophysics Division, Postbus 800, 9700 AV Groningen, the Netherlands\\
$^{10}$ Department of Microelectronics, Faculty of Electrical Engineering, Mathematics and Computer Science, Delft University of Technology, 2628 CD Delft, The Netherlands.\\
$^{11}$  Department of Quantum Nanoscience, Kavli Institute of Nanoscience, Delft University of Technology, Lorentzweg 1, 2628 CJ, Delft, the Netherlands.\\
$^{12}$ Netherlands Institute for Radio Astronomy (ASTRON), Postbus 2, 7990 AA, Dwingeloo, The Netherlands\\
$^{13}$ Astronomical Observatory Institute, Faculty of Physics, A. Mickiewicz University, Sloneczna 36, 60-286, Poznan, Poland\\
$^{14}$ Jet Propulsion Laboratory, California Institute of Technology, 4800 Oak Grove Drive, Pasadena, CA 91109, USA\\
$^{15}$ Netherlands Research School for Astronomy, P.O. Box 9513, 2300 RA Leiden, the Netherlands\\
}
\date{Received Jan 1, 2013}

 \abstract
   {Protostars of intermediate mass provide a bridge between theories of low- and high-mass star formation. Molecular outflows emerging 
   from such sources can be used to determine the influence of fragmentation and multiplicity on protostellar evolution through the apparent 
   correlation of outflow forces of intermediate mass protostars with
   the total luminosity instead of the individual luminosity.}
   {The aim of this paper is to derive outflow forces from outflows of six intermediate mass protostellar regions and 
   validate the apparent correlation between total luminosity and outflow force seen in earlier work, as well as remove uncertainties caused by different methodology. }
   {By comparing CO 6--5 observations obtained with APEX with non-LTE radiative transfer model predictions, 
   optical depths, temperatures, densities of the gas of the molecular outflows are 
   derived. Outflow forces, dynamical timescales and kinetic luminosities are subsequently calculated. }
   {Outflow parameters, including the forces, were derived for all sources. Temperatures in excess of 50 K were found for all flows, in line with recent low-mass results. However, comparison with other studies could not corroborate conclusions from earlier work on intermediate mass protostars which hypothesized that fragmentation enhances outflow forces in clustered intermediate mass star formation. Any enhancement in comparison with the classical relation between outflow force and luminosity can be attributed the use of a higher excitation line and improvement in methods; They are in line with results from low-mass protostars using similar techniques.
   }
   {The role of fragmentation on outflows is an important ingredient to understand clustered star formation and the link between low and high-mass star formation. However, detailed information on spatial scales of a few 100 AU, covering all individual members is needed to make the necessary progress.
    }


\maketitle

%

\def\placeFigureSingleCOSixFive{
\begin{figure}[!tp]
\centering
\includegraphics[width=8cm]{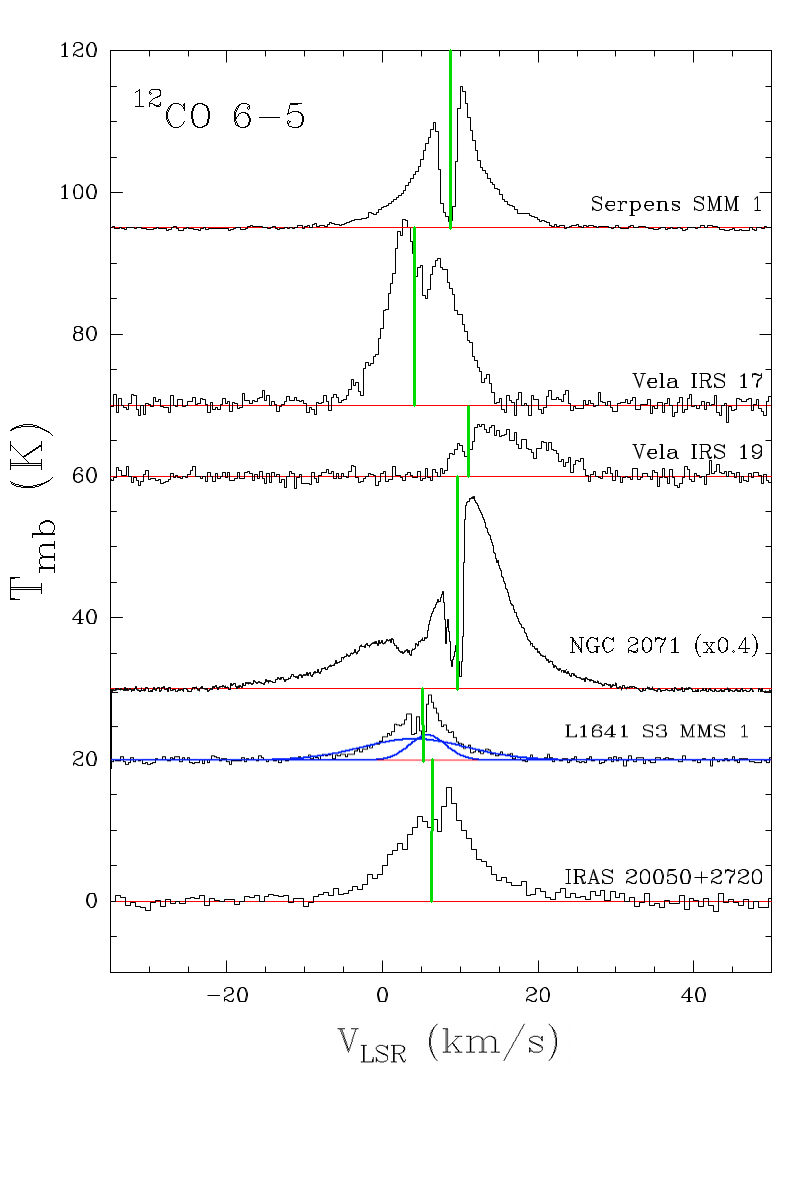}
\caption{CO 6--5 spectra taken at the central position. The baseline is shown in red. The line profile for L1641 S3 MMS 1 shows an example 
of the gaussian component fit (overplot in blue). The green lines show the source velocity, also listed in Table 1. }
\label{fig:CO65central}
\end{figure}
}

\def\placeFigureSingleCOSevenSix{
\begin{figure}[!tp]
\centering
\includegraphics[width=8cm]{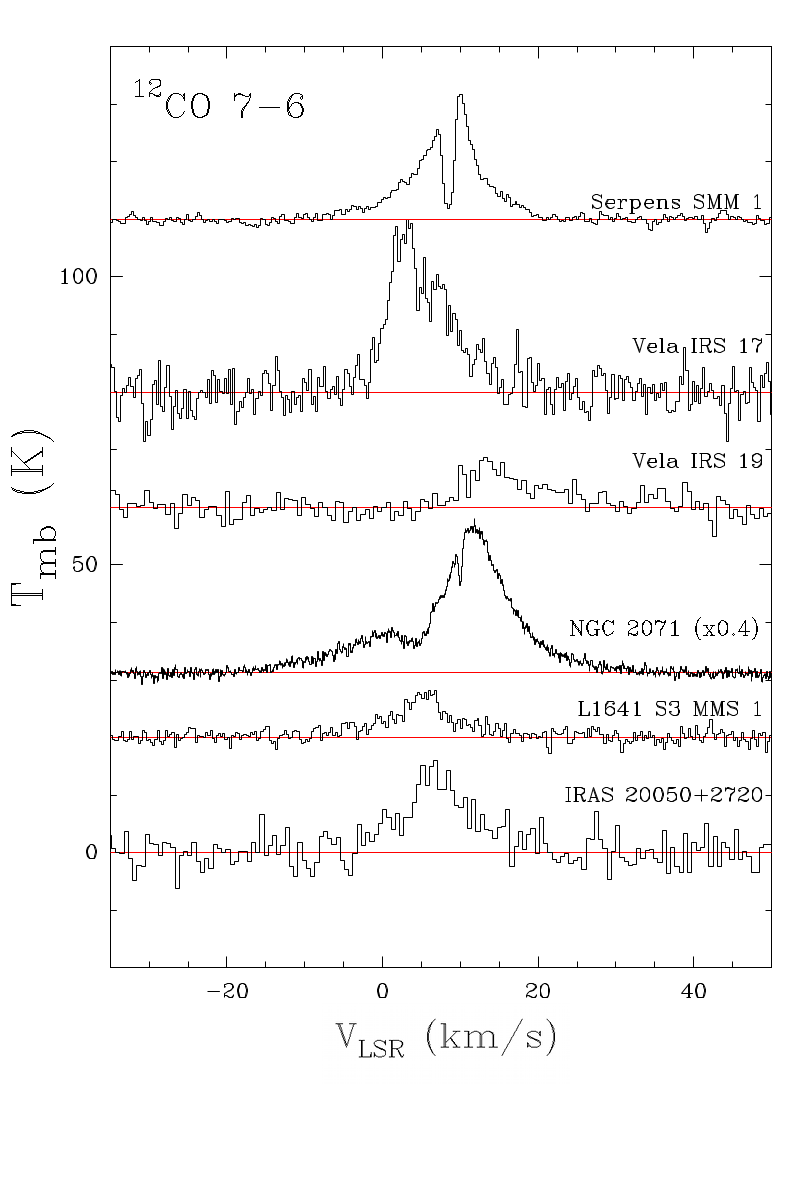}
\caption{CO 7--6 spectra taken at the central position. The baseline is shown in red.}
\label{fig:CO76central}
\end{figure}
}

\def\placeFigureSingle13CO{
\begin{figure}[!tp]
\centering
\includegraphics[width=8cm]{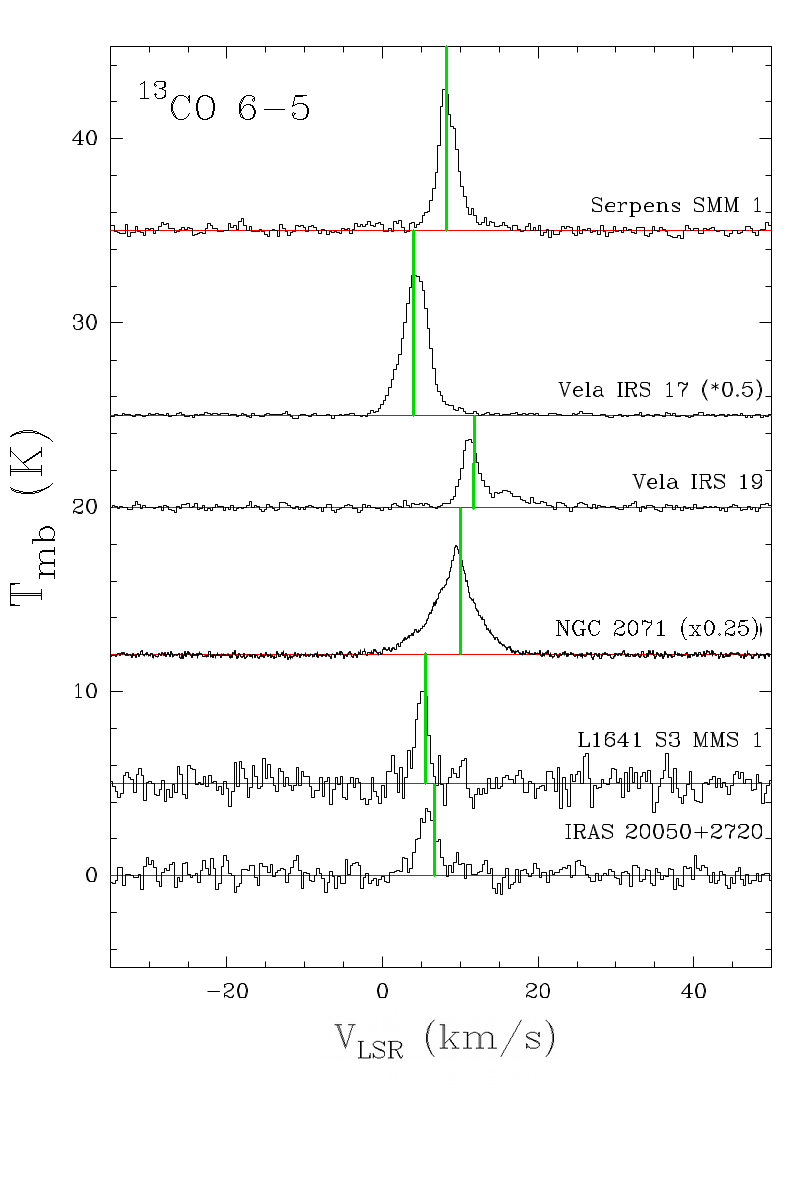}
\caption{$^{13}$CO 6--5 spectra taken at the central position. The baseline is shown in red. The green lines show the source velocity, also listed in Table 1.  }
\label{fig:13COcentral}
\end{figure}
}

\def\placeFigureSingleCarbon{
\begin{figure}[!tp]
\centering
\includegraphics[width=8cm]{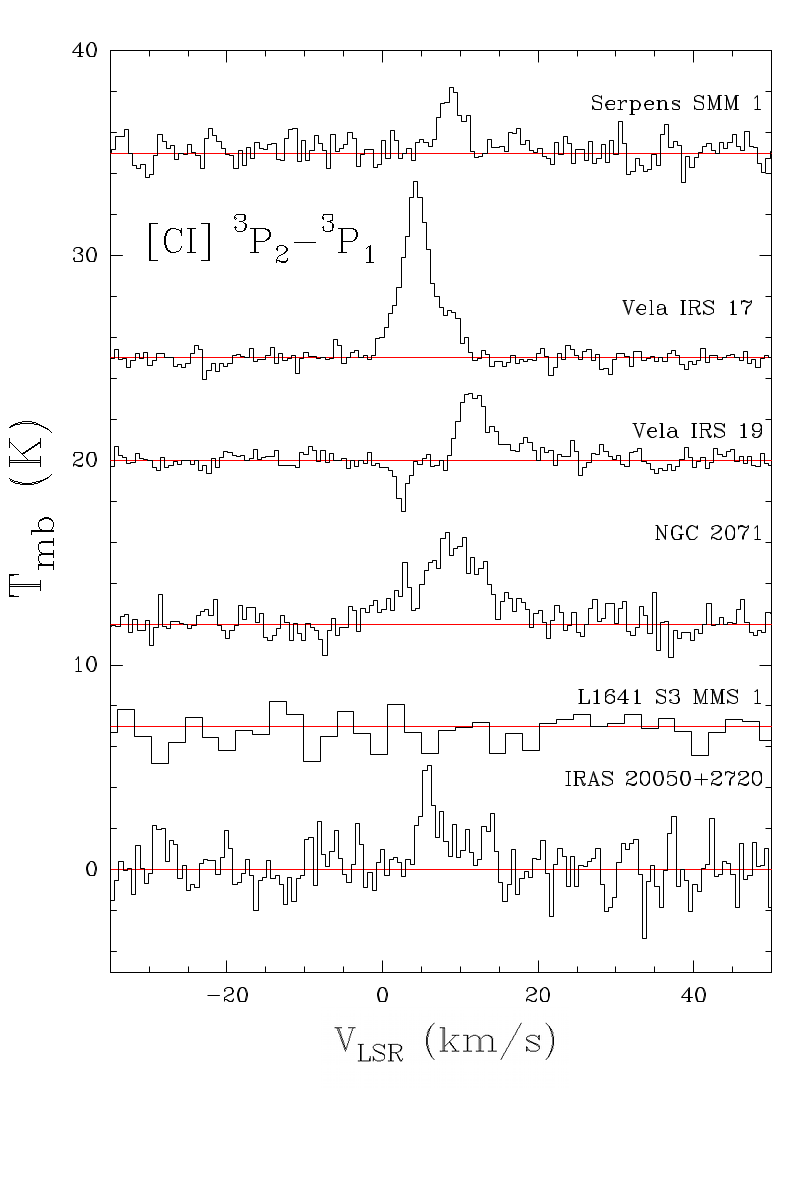}
\caption{[CI] $^3$P$_2$--$^3$P$_1$ spectra taken at the central position. 
The baseline is shown in red. The absorption feature at $\sim$0 km s$^{-1}$ in Vela IRS 19 is from the off-position but does not affect
the main line emission. }
\label{fig:carboncentral}
\end{figure}
}

\def\placeCOSixFiveMaps{
\begin{figure*}[!tp]
\centering
\includegraphics[width=8.5cm]{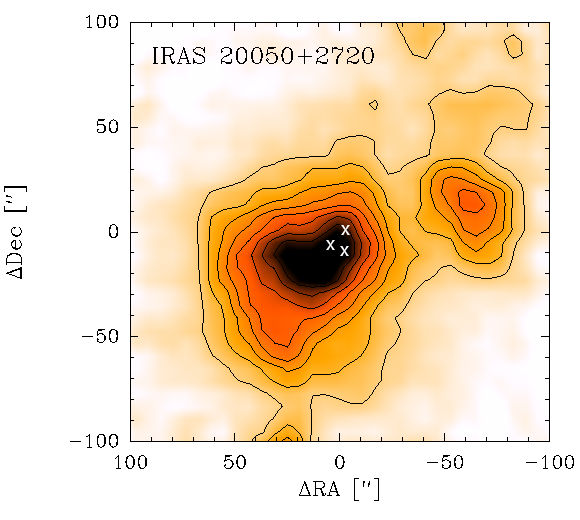}
\includegraphics[width=8.5cm]{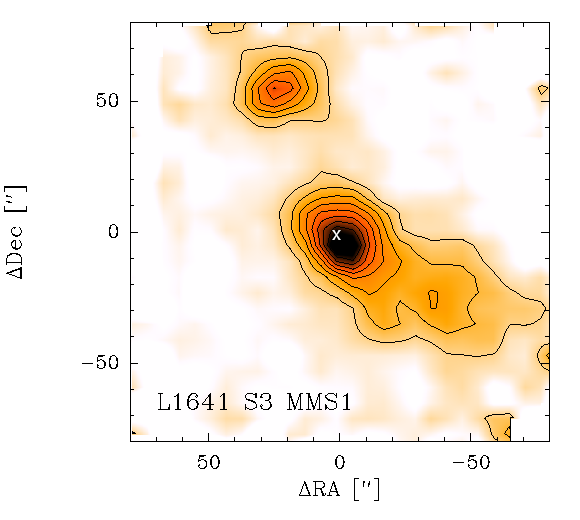}
\includegraphics[width=8.5cm]{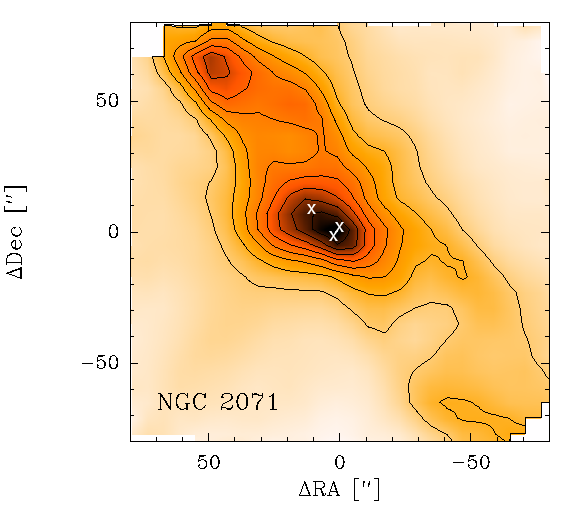}
\includegraphics[width=8.5cm]{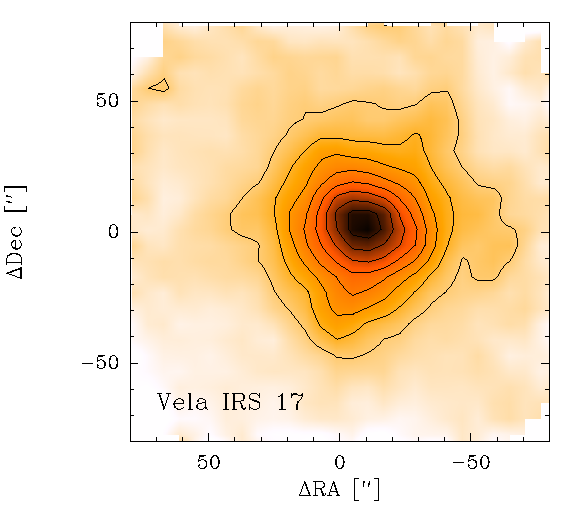}
\includegraphics[width=8.5cm]{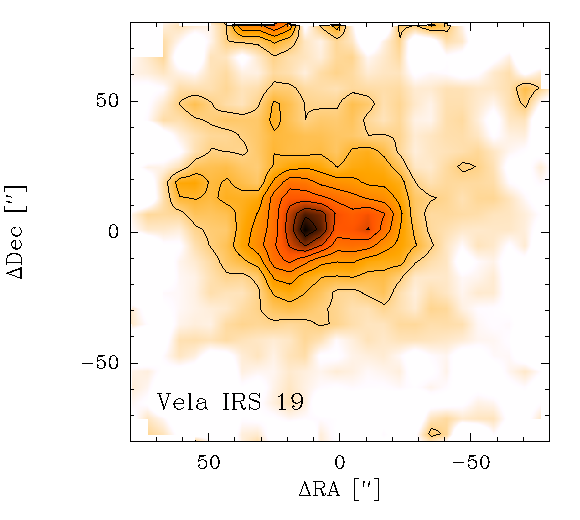}
\includegraphics[width=8.5cm]{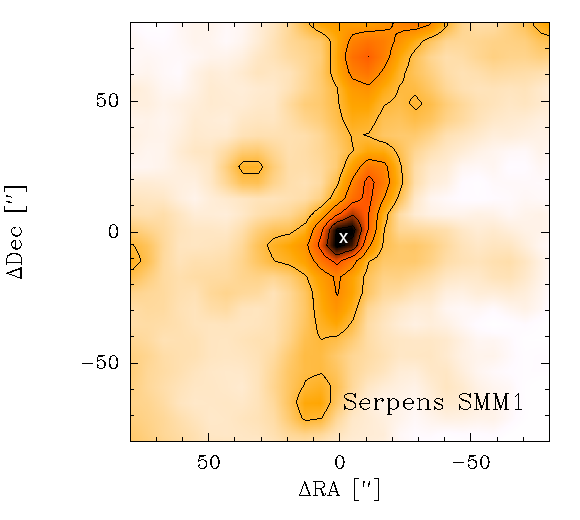}

\caption{Integrated intensity maps of CO 6--5. The colorscale is normalized to the central integrated intensity in K km s$^{-1}$ given in Table 2, with the white
background corresponding to the 3$\sigma$ noise level. This was done to emphasise larger-scale CO 6--5 emission and its
strength relative to the central position. Contour levels are given in 20$\%$, 30$\%$,..., 80$\%$, 90$\%$ 
of the central integrated intensity, which can be found in Table 2. Where known, locations of 
(sub)millimeter interferometry sources are shown with `x', except for the Vela sources.}
\label{fig:CO65maps}
\end{figure*}
}

\def\placeCOthirteenMaps{
\begin{figure*}[!tp]
\centering
\includegraphics[width=8.5cm]{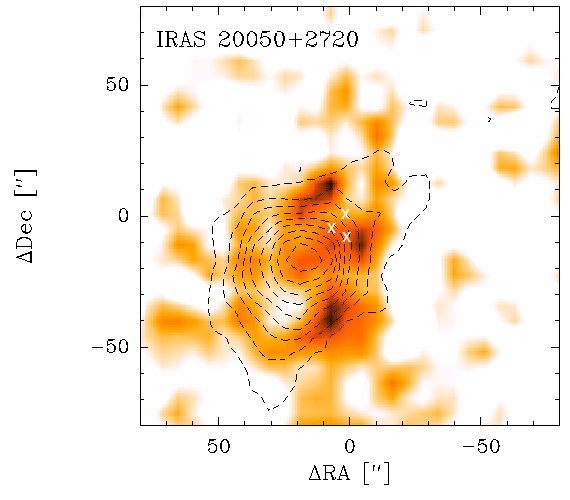}
\includegraphics[width=8.5cm]{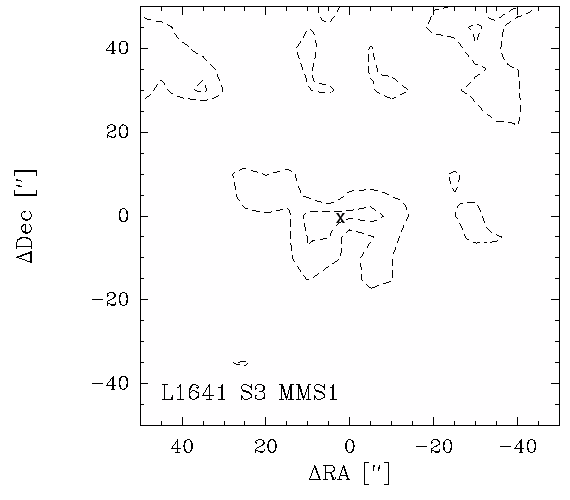}
\includegraphics[width=8.5cm]{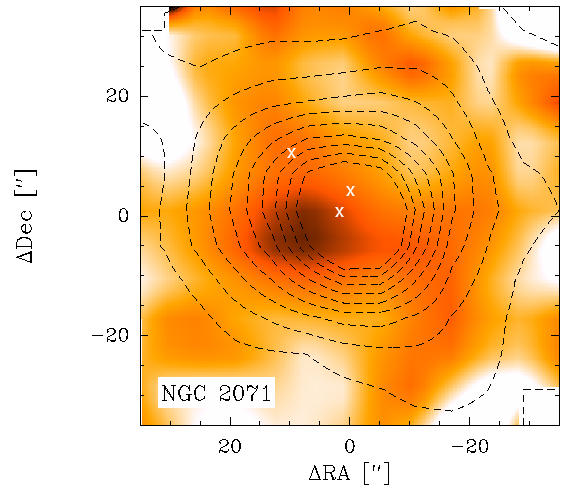}
\includegraphics[width=8.5cm]{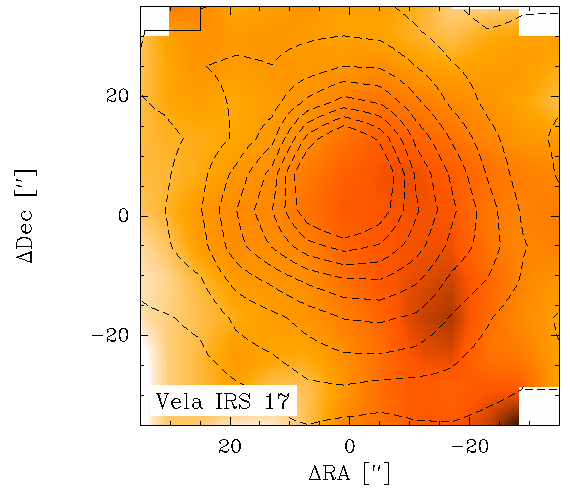}
\includegraphics[width=8.5cm]{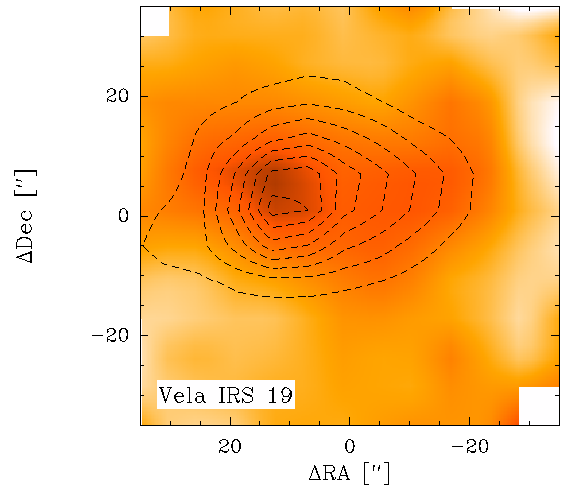}
\includegraphics[width=8.5cm]{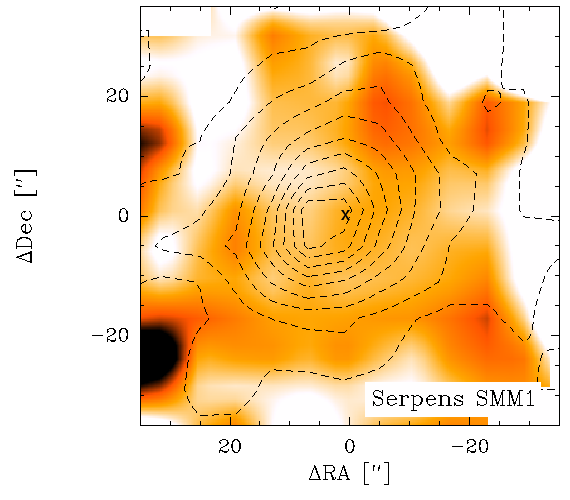}
\caption{Integrated intensity of $^{13}$CO 6--5 (\textit{contours}) overplotted on the integrated intensity
of [CI] (\textit{colorscale}). Both distributions are normalized to the peak integrated intensity 
of that particular line. For $^{13}$CO the contours are in levels of  10$\%$, 20$\%$, ..., 80$\%$, 90$\%$ w.r.t. to this peak intensity, 
with the lowest contour higher than 3 times the noise level in Table 2. The [CI] emission scales between 3 times the noise 
level and the highest intensity in the map, which can be found in Table 2. Where known, locations of 
(sub)millimeter interferometry sources are shown with `x', except for the Vela sources.}
\label{fig:13COmaps}
\end{figure*}
}

\def\placeFlowMaps{
\begin{figure*}[!tp]
\centering
\includegraphics[width=8.3cm]{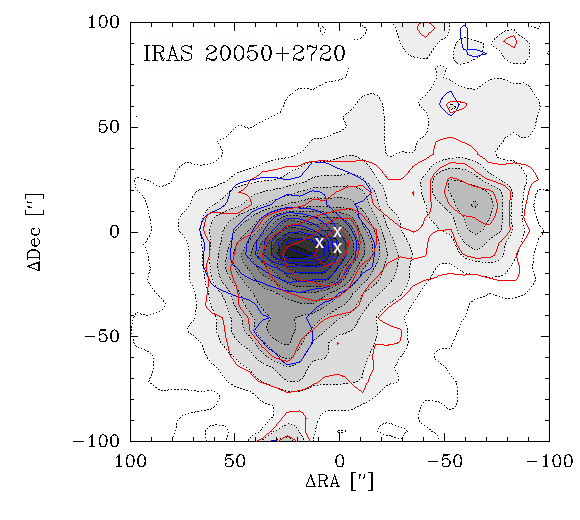}
\includegraphics[width=8.3cm]{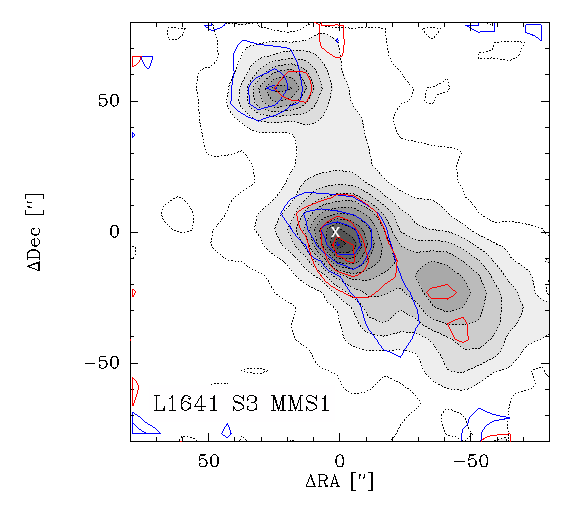}
\includegraphics[width=8.3cm]{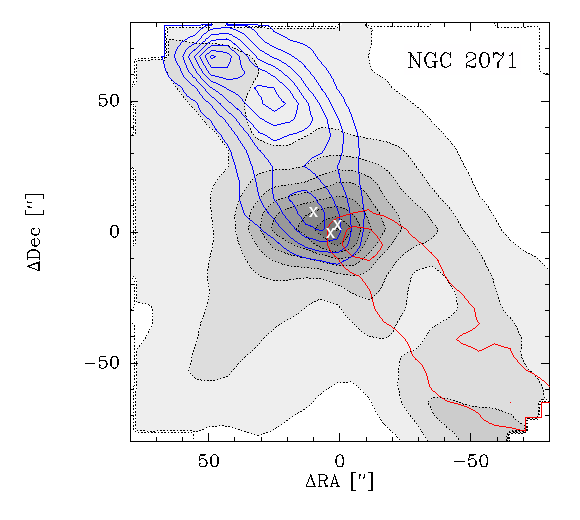}
\includegraphics[width=8.3cm]{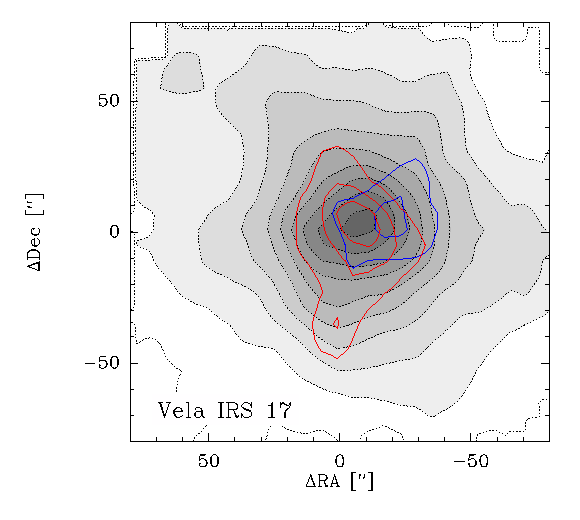}
\includegraphics[width=8.3cm]{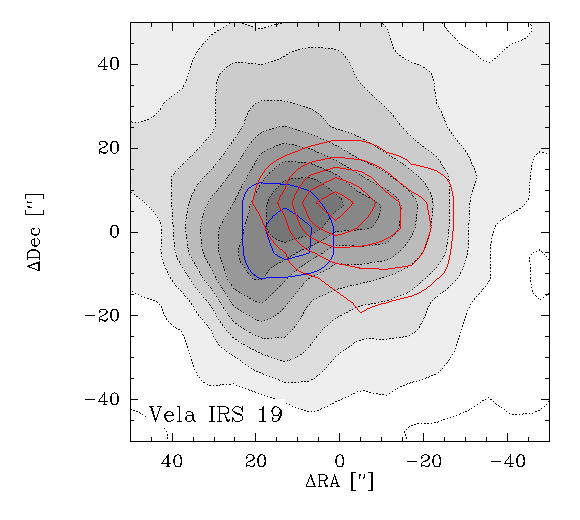}
\includegraphics[width=8.3cm]{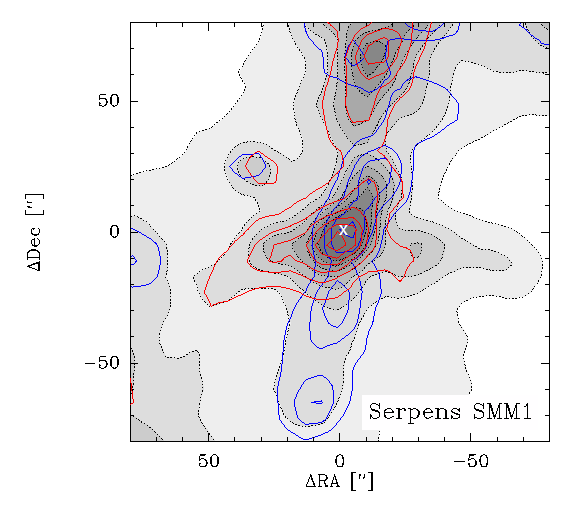}
\caption{Maps showing the CO $J$=6--5 integrated emission in  -20 to -4 (\textit{blue}) and +4 to +20 (\textit{red}) 
km s$^{-1}$ bins to visualize the outflowing gas, overplotted on the integrated emission in a -3 to +3 
km s$^{-1}$ bin (\textit{dotted line and grayscale}), representing the quiescent emission. 
All velocities are with respect to the individual source velocity (see Table 1). 
NGC 2071 was characterized by defining outflow bins at -40 to -10 and +10 
to +40 km s$^{-1}$ instead of the bins above.
All contours, including those of the outflowing gas are normalized towards the peak intensity of the \textit{quiescent} gas component 
at the central position ($T_{\rm{peak}}$ in Table 2). 
Levels are in turn given in 10$\%$, 20$\%$, ..., 80$\%$, 90$\%$ w.r.t. to this peak intensity. Where known, locations of 
(sub)millimeter interferometry sources are shown with `x'.}
\label{fig:flowmaps}
\end{figure*}
}

\def\placespecN{
\begin{figure*}[!tp]
\centering
\includegraphics[width=15cm]{Pictures/N2071_spectra_CO.png}
\caption{Spectral line profiles of CO 6--5 for NGC 2071 for the inner 2$'$ by 2$'$. The 
velocity scale per spectra ranges from -40 to +55 km s$^{-1}$. The y scale ranges from -3 to +15 Kelvin in $T_{\rm{mb}}$,
and emphasize the `narrow' CO 6--5 emission across the core in all directions. }
\label{fig:2071spec}
\end{figure*}
}

\def\placespecI{
\begin{figure}[!tp]
\centering
\includegraphics[width=8cm]{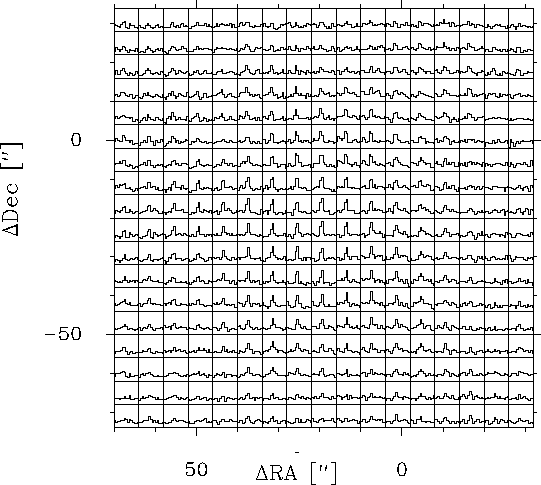}
\caption{Spectral line profiles of [CI] of IRAS 20050+2720 for the inner 2$'$ by 2$'$, centered on the main core. The 
x scale per spectra ranges from -10 to +30 km s$^{-1}$. The y scale, ranging in $T_{\rm{mb}}$ from -2 to +7 Kelvin,
has been chosen to emphasize the `narrow' [CI] emission across the core in all directions. }
\label{fig:20050spec}
\end{figure}
}

\def\placeiras{
\begin{figure*}[!tp]
\centering
\includegraphics[width=17cm]{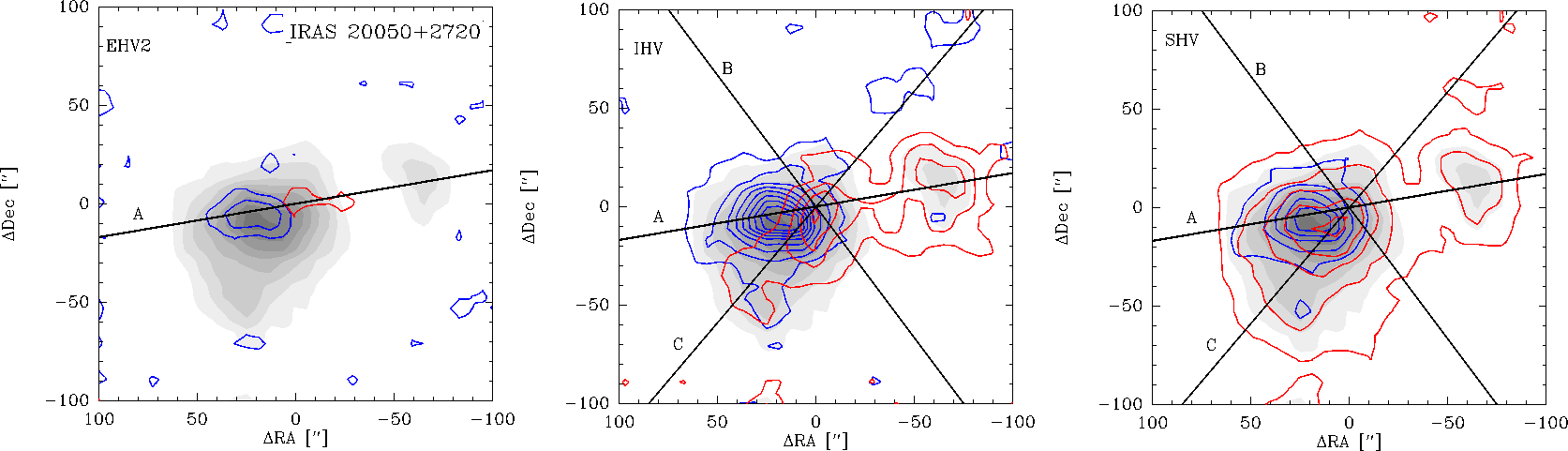}
\caption{CO 6--5 Emission of IRAS 20050+2720 in the same velocity bins defined by \cite{1995ApJ...445L..51B} with the 
A, B and C flows identified by black lines: \textit{Left} EHV2 (-39 to -14 and 26 to 51 km s$^{-1}$), \textit{middle} 
IHV (-14 to -4 and 16 to 26 km s$^{-1}$ and \textit{right} SHV (-4 to 1 and 11 to 16 km s$^{-1}$). The `narrow' emission 
in a bin of  $\pm$ 2 km s$^{-1}$ bin around 6.4 km s$^{-1}$ is shown in greyscale. No detections were found for 
EHV1 (-89 to -39 and 51 to 101 km s$^{-1}$). All red and blue contour lines are drawn at 3, 6, 9, 12$\sigma$ ...}
\label{fig:EHV20050}
\end{figure*}
}

\def\placeehvser{
\begin{figure*}[!tp]
\centering
\includegraphics[width=12cm]{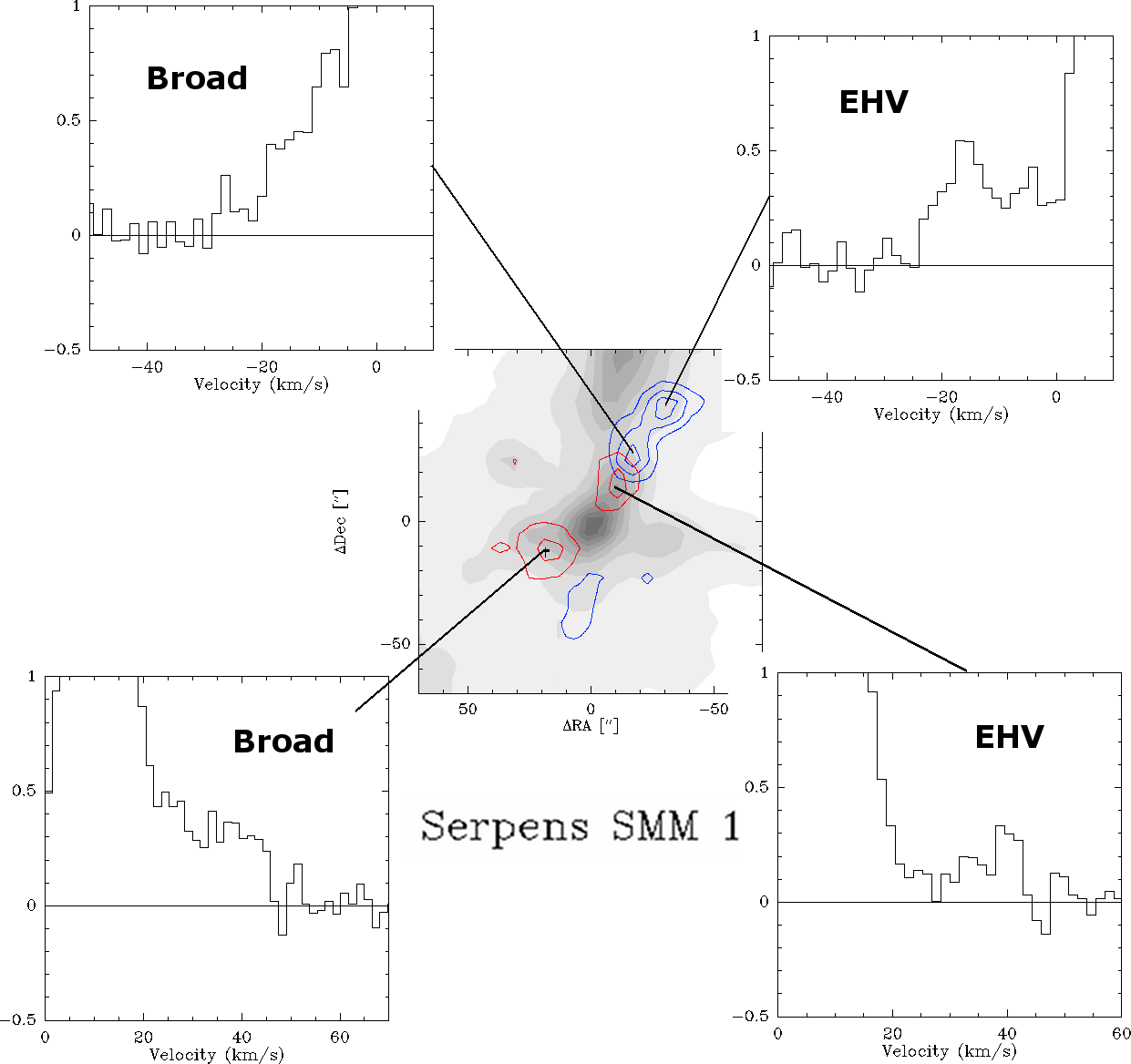}
\caption{Difference between the CO 6--5 EHV and `broad' components in Serpens SMM1. The insets are the profiles at 
EHV speeds. In the center images, the red and blue contours are the integrated intensities around EHV velocities 
at 3, 6, 9 $\sigma$. 
The greyscale is the CO emission near the source velocity.}
\label{fig:EHVS}
\end{figure*}
}

\def\placeradex{
\begin{figure}[!tp]
\centering
\includegraphics[width=8cm]{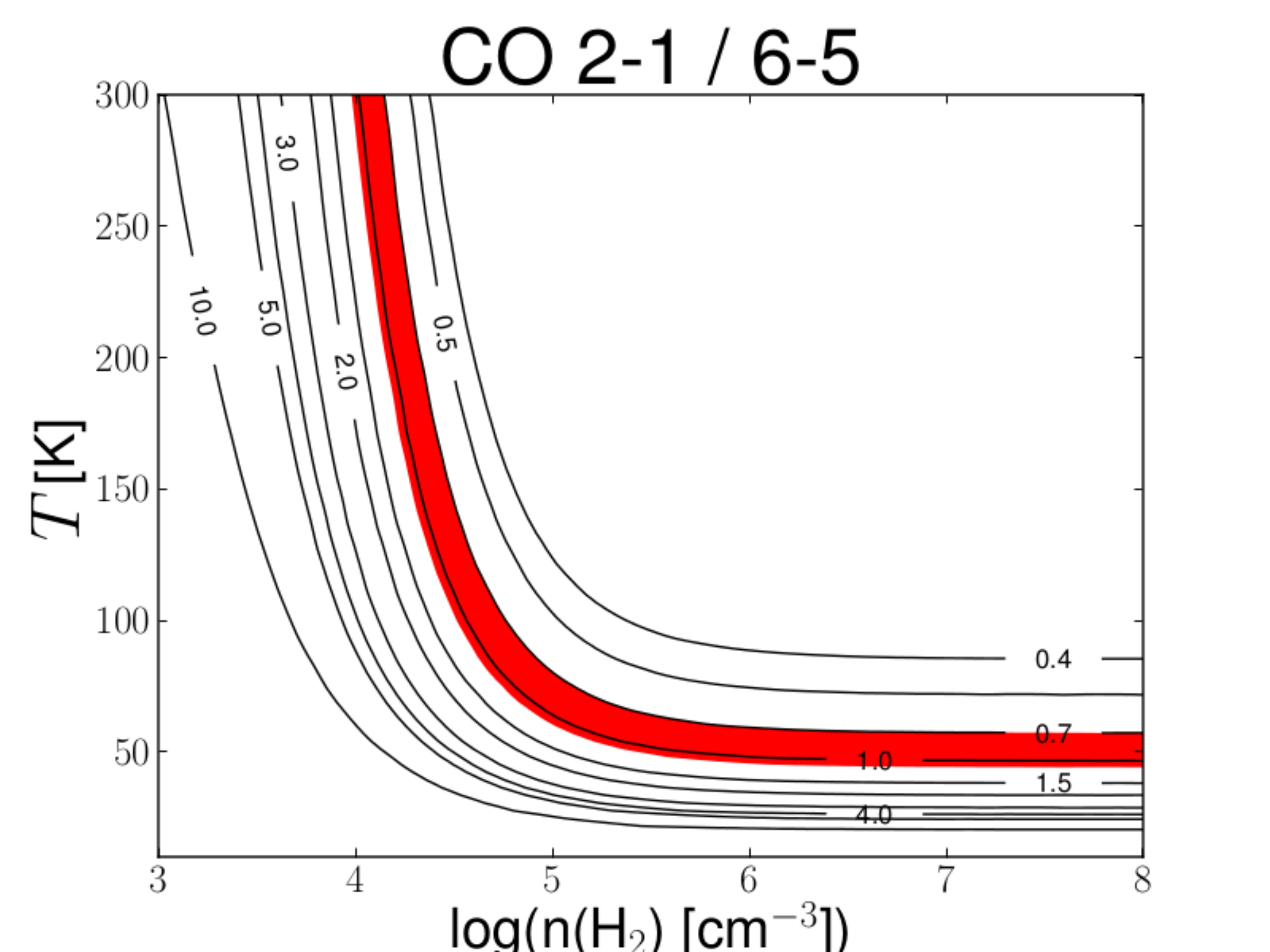}
\includegraphics[width=8cm]{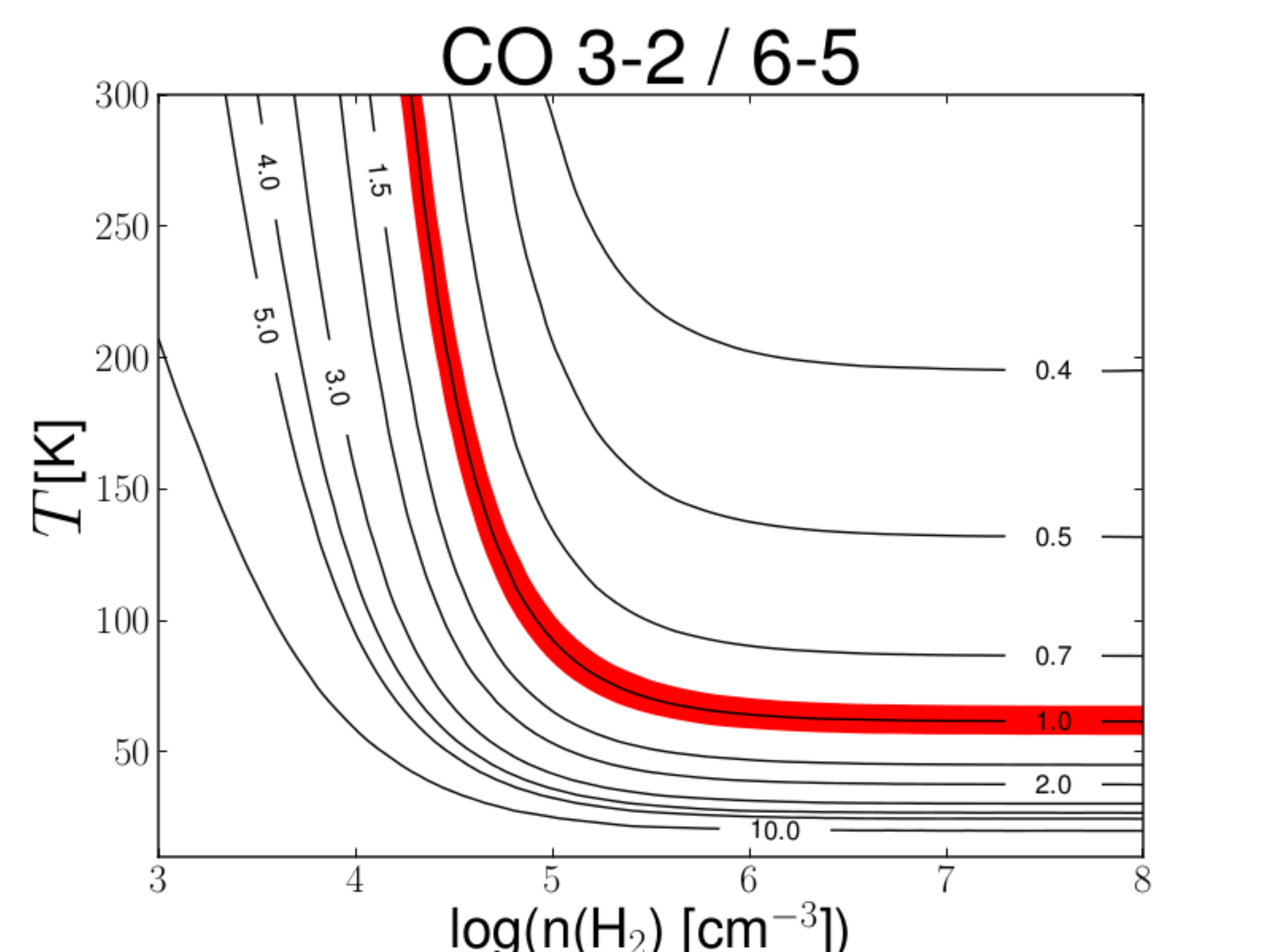}
\includegraphics[width=8cm]{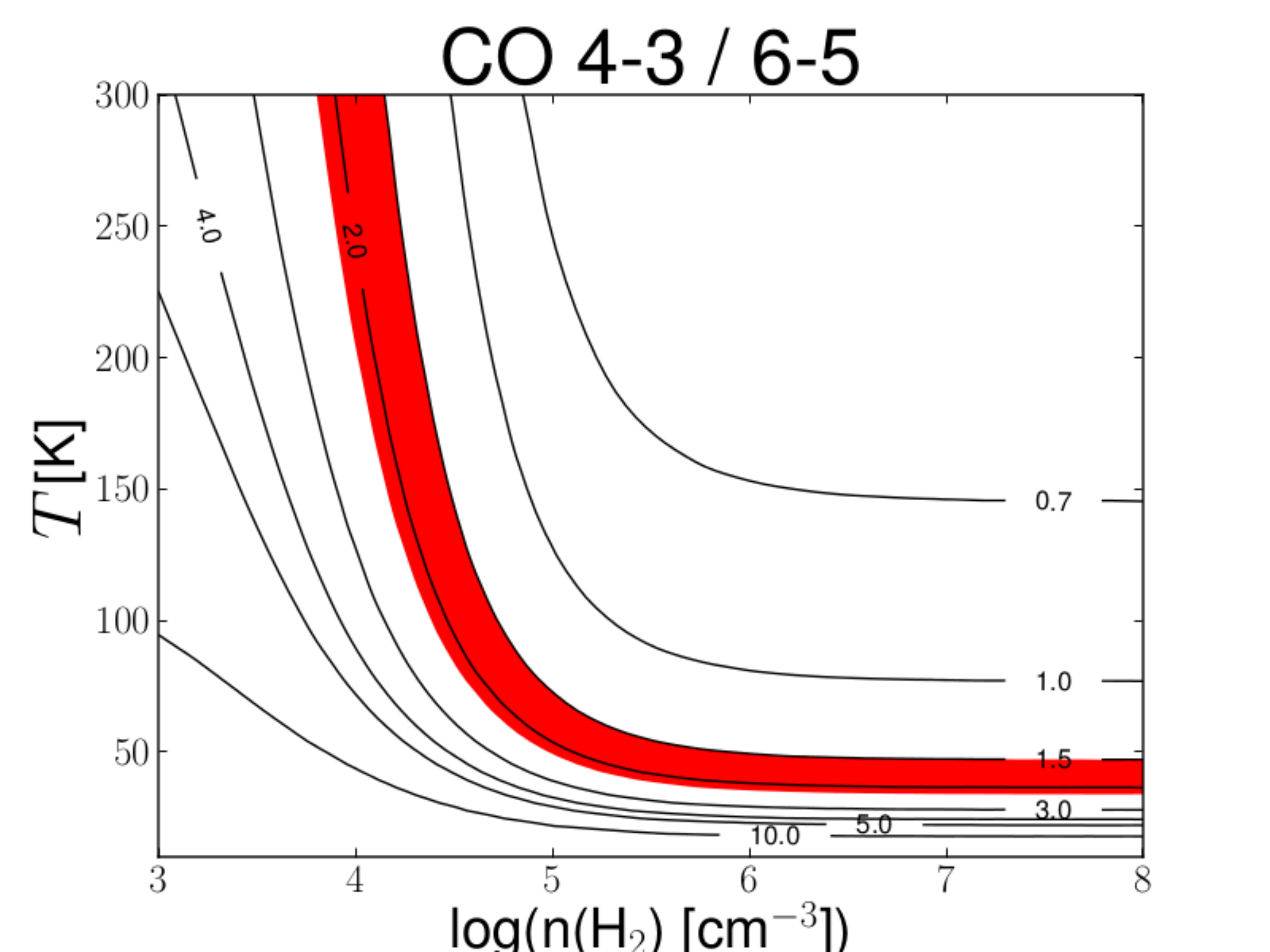}
\caption{RADEX diagnostic plots of CO 2--1 (\textit{top}), 3--2 (\textit{middle}) and 4--3 (\textit{bottom}). 
The derived line ratios of the blue flow of Serpens SMM1 are highlighted in red to serve as an illustration for the sub-thermal and thermal excitation 
scenarios. It is evident that temperatures below 50 K are excluded in both limits as ratios of CO 2--1 and 3--2 over 6--5 do not agree with those of 4--3.}
\label{fig:radex}
\end{figure}
}

\def\placefigwidth{
\begin{figure}[!tp]
\centering
\includegraphics[width=9cm]{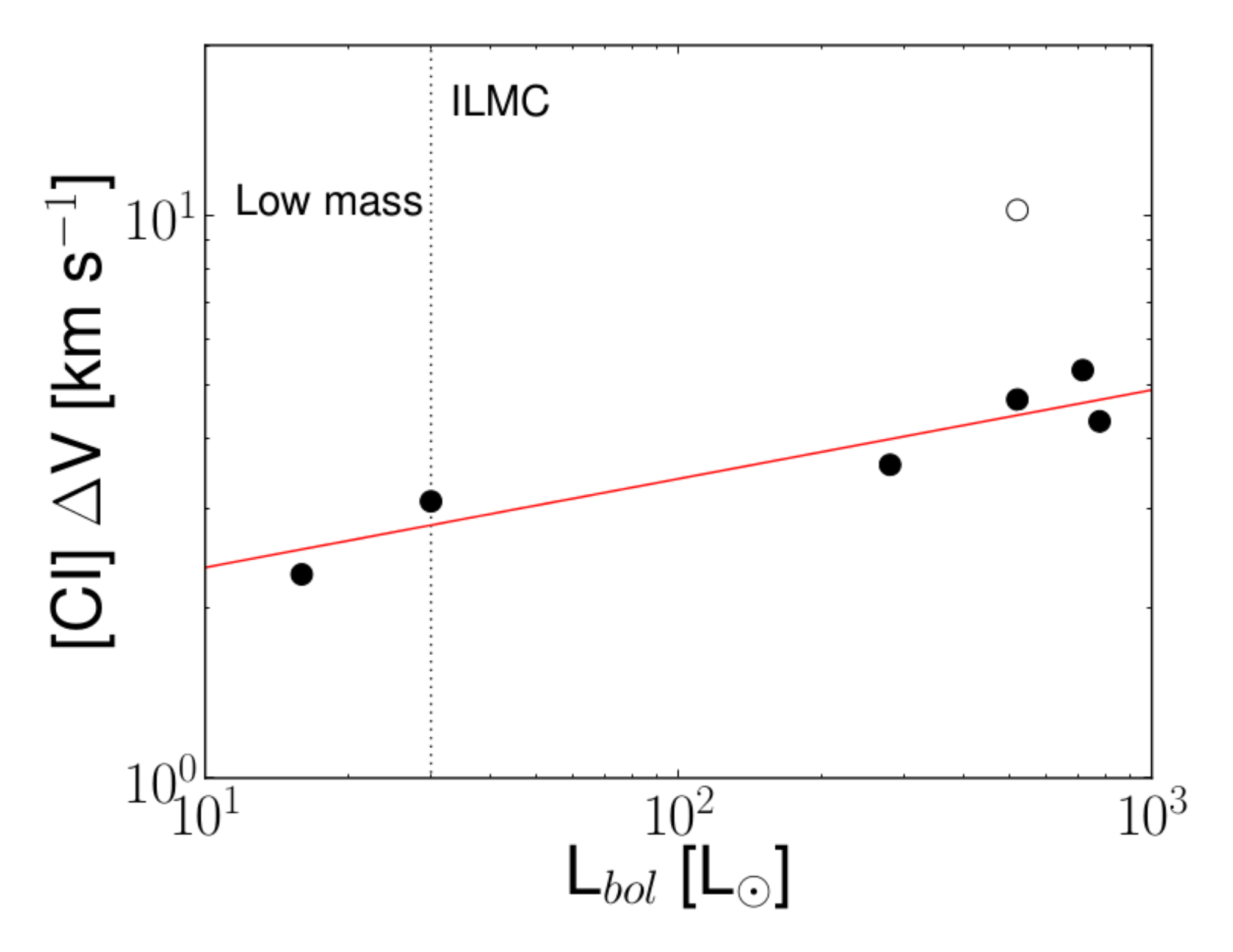}
\caption{Line width of the [CI] line versus total bolometric luminosity. The result from HH~46 \citep{2009A&A...501..633V} is also included. 
As an outlier, the on-source measurement 
of NGC 2071 is shown with an open circle. A filled circle is a [CI] measurement 20$''$ away from the NGC 2071 flow. This measurement is likely affected by the outflow. 
 The 
fit without NGC 2071 on-source is shown in a red line.}
\label{fig:width}
\end{figure}
}

\def\placeLineLum{
\begin{figure}[!tp]
\centering
\includegraphics[width=9cm]{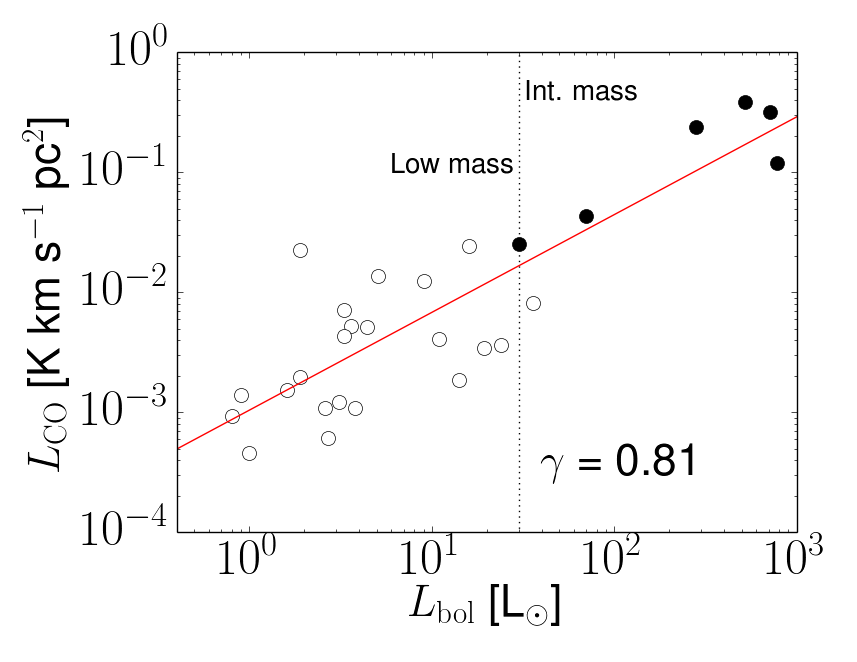}
\includegraphics[width=9cm]{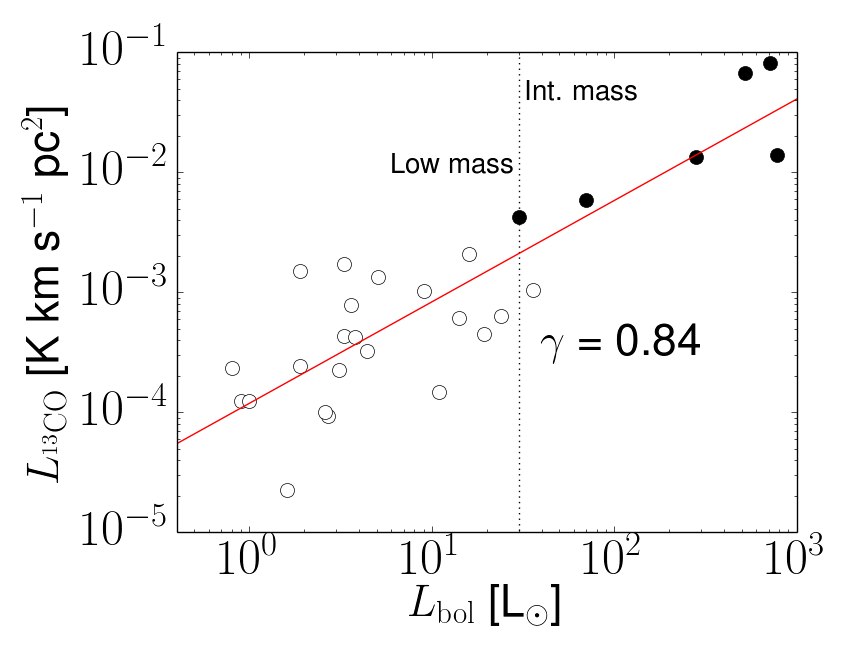}
\includegraphics[width=9cm]{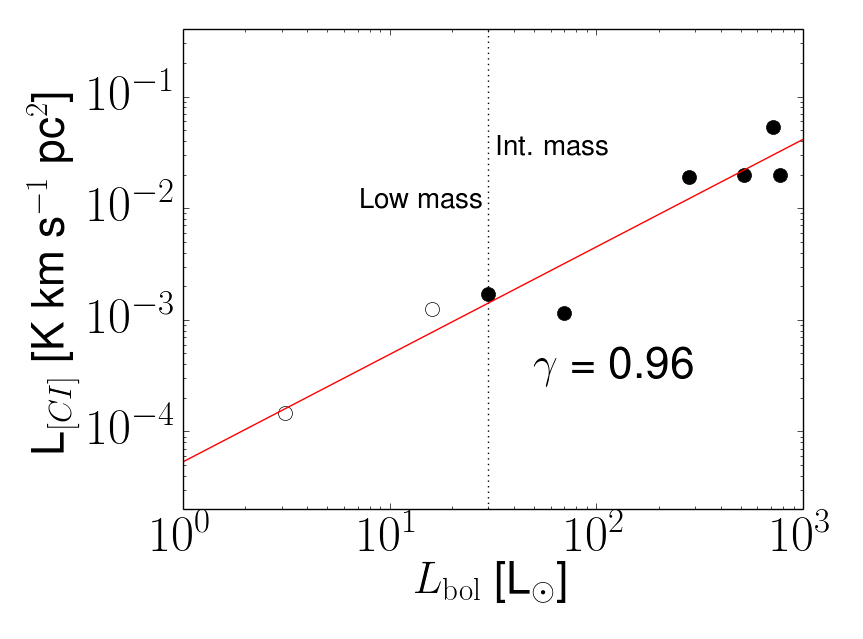}
\caption{{\it Top:} CO 6--5 line luminosity vs $L_{\rm{bol}}$, {\it Middle:} $^{13}$CO 6--5 line luminosity vs 
$L_{\rm{bol}}$, {\it Bottom:} [CI] $^3$P$_2$--$^3$P$_1$ line luminosity vs $L_{\rm{bol}}$. Fits to the line 
luminosities are shown in red, with the slope labelled '$\rm{\gamma}$'. The slopes are 
within the error bars of \cite{2013A&A...553A.125S}. Open symbols are the low-mass sample from \cite{2013A&A...556A..89Y}. Note
that no [CI] was reported in \cite{2013A&A...556A..89Y}. The [CI] detections of \cite{2009A&A...507.1425V} are included. }
\label{fig:Linelum}
\end{figure}
}

\def\placeFigureFcoLum{
\begin{figure*}[!tp]
\centering
\includegraphics[width=14cm]{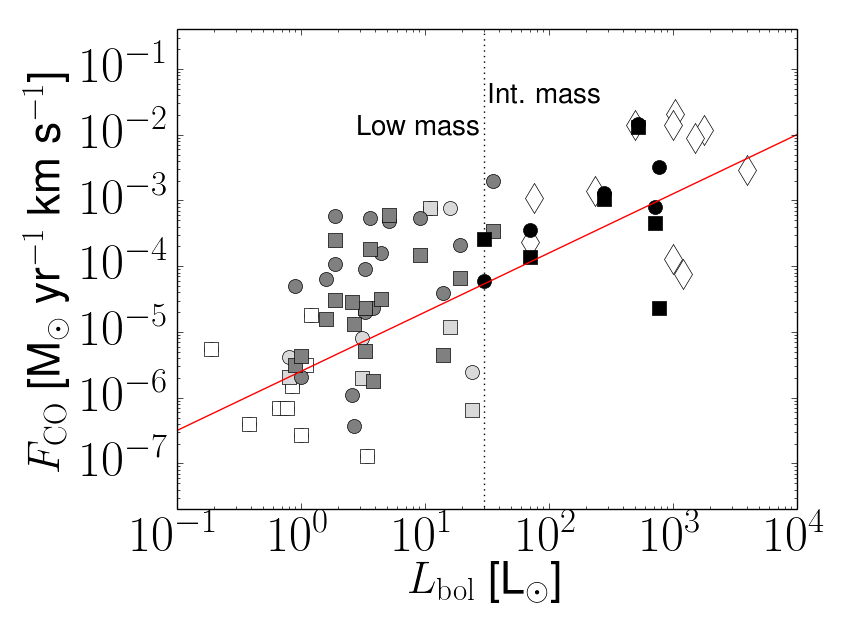}
\caption{Luminosity versus outflow force of the blue (circle) and red (square) lobes. Black symbols
are the derived values for the intermediate mass protostars, dark grey symbols are values from \citet{2015A&A...576A.109Y},
light grey symbols are values from \cite{2009A&A...507.1425V} and white symbols from \citet{2013A&A...556A..76V}. 
All values except the ones from \cite{2013A&A...556A..76V} are derived using CO 6--5.
White diamonds are the values from B08.  The red line 
represents the relation proposed by \cite{1996A&A...311..858B}.}
\label{fig:fcolum}
\end{figure*}
}

\def\placeFigureFcoMenv{
\begin{figure*}[!tp]
\centering
\includegraphics[width=14cm]{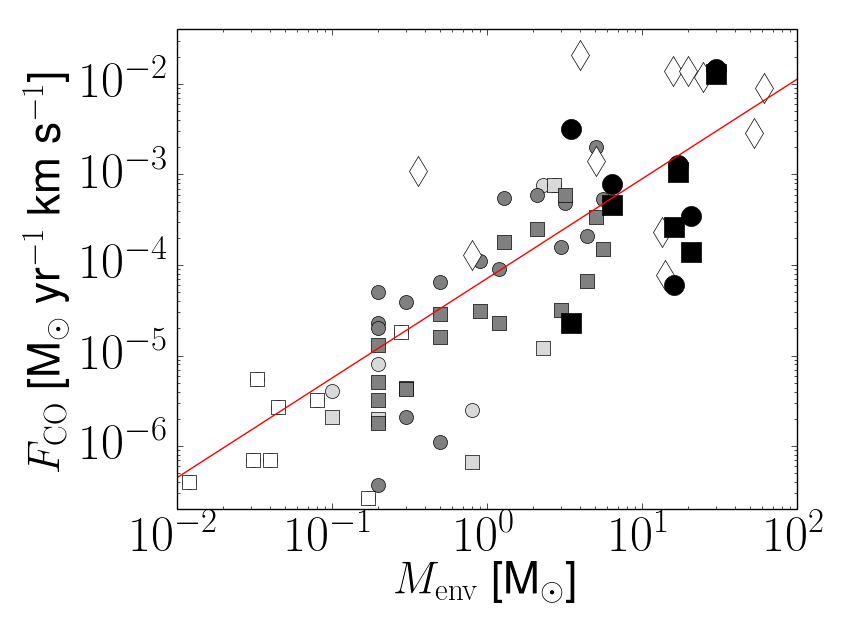}
\caption{Envelope masses versus outflow force of the blue (circle) and red (square) lobes. 
Black symbols
are the derived values for the intermediate mass protostars, dark grey symbols are values from \citet{2015A&A...576A.109Y},
light grey symbols are values from \cite{2009A&A...507.1425V} and white symbols from \cite{2013A&A...556A..76V}.
White diamonds are the values from B08. 
The red line represents the relation proposed by \cite{1996A&A...311..858B}. }
\label{fig:fcomenv}
\end{figure*}
}

%

\def\placeTableSources{
\begin{table*}[!tp]
\caption{Sample of studied IM protostars. The Right Ascension and Declination are the previously derived and/or estimated centers of gravity.}
\begin{center}
\begin{tabular}{l c c c c c c c c c c}
\hline \hline
Source & R.A. & Decl. & $L_{\rm{bol}}$ & Dist. & Mass & $V_{\rm{LSR}}$ & \# &Ref.$^1$ \\ 
& (hms [J2000]) & (dms [J2000]) & (L$_\odot$) & (pc) & (M$_\odot$) & (km s$^{-1}$) & members\\ \hline
NGC 2071 & 05:47:04.4 & +00:21:49.0 & 520/40$^2$ & 422 & 30 & 9.6 & 5 & 1,2\\
L1641 S3 MMS 1 &  05:39:55.9 & -07:30:28.0 &  70/250$^2$ & 465 & 20.9 & 5.3 & 1 & 1,2,3,4\\
Vela IRS 17 & 08:46:34.7 & -43:54:30.5 & 715 & 700 & 6.4 &3.9 & 3+$^3$ & 5,6,7,8 \\
Vela IRS 19 & 08:48:48.5 & -43:32:29.0 & 776 & 700 & 3.5 &12.2 & 3+$^3$ & 5,6,8\\
Serpens SMM1 &  18:29:49.8& +01:15:20.5 & 30 & 250&16.1 & 8.5 & 1 & 9,10,11,12,13\\ 
IRAS 20050+2720 & 20 07 05.8 & +27 29:00.0 & 280 & 700& 17.1 &6.4 & 4+ & 14,15,16,17,18,19\\ 
\hline
\end{tabular}
\end{center}
\tiny $^1$ 
References beyond \cite{2011PASP..123..138V} - 1: \cite{2005A&A...430..523W}, 2: \cite{2012ApJ...751..137V}, 
3: \cite{2000A&A...355..639S}, 4:  \cite{1991ApJ...376..618M} as FIRSSE 101, 
5:\cite{1992A&A...265..577L}, 6: \cite{1988AJ.....96..988S}, 7:\cite{2005A&A...433..941G}, 8 :\cite{1999A&AS..140..177W}, 
9: \cite{2010A&A...516A..57K}, 10:  \cite{2012A&A...548A..77G}, 11: \cite{1999ApJ...513..350H}, 12: \cite{1995A&A...298..594W}, 
13: \cite{2012A&A...542A...8K}, 14: \cite{2005ApJS..156..169F}, 
15: \cite{1995ApJ...445L..51B}, 16: \cite{2005ApJ...625..864Z}, 17: \cite{2001A&A...369..155C}, 
18:\cite{2009A&A...507..369W}, 19: B08.\\
$^2$ Second luminosity numbers from \citet{2012ApJ...751..137V}, equalling the sum of individual luminosities.\\
$^3$ Estimated from the infrared results \citep{2005A&A...433..941G}.
\label{tab:sources}
\end{table*}
}

\def\placeTableTotal{
\begin{table}[!tp]
\caption{Noise levels, integrated and peak intensities at the central position. Temperatures are in given in main beam temperature units.}
\begin{center}
\begin{tabular}{l ccc}
\hline \hline
Source & \multicolumn{3}{c}{CO 6--5}  \\ 
& $\int T_{\rm{MB}} dV$ & $T_{\rm{peak}}$& RMS$^2$ \\
& [K km s$^{-1}$] & [K] & [K] \\ \hline

NGC 2071        	& 819.7  	& 64.0 &0.27 \\
L1641 S3 MMS 1 	& 75.0   	& 8.8  &0.30\\
Vela IRS 17	   		& 241.7  	& 25.6 & 0.9 \\
Vela IRS 19 	   	& 91.2		& 7.1  & 0.78 \\
IRAS 20050+2720 	& 183.4	 	& 16.2 & 0.60 \\
Serpens SMM1 		& 151.3 	& 19.8 & 0.20\\ \hline
& \multicolumn{3}{c}{$^{13}$CO 6--5}\\
& $\int T_{\rm{MB}} dV$ & $T_{\rm{peak}}$ & RMS$^1$  \\
& [K km s$^{-1}$] & [K] & [K] \\ \hline
NGC 2071        	& 141.1& 22.8 & 0.40 \\
L1641 S3 MMS 1 	& 10.2 &  5.0 & 0.60\\
Vela IRS 17	   		& 63.1 & 14.8 & 0.12 \\
Vela IRS 19 	   	& 10.8 & 3.6 & 0.12\\
IRAS 20050+2720 	& 10.3 & 3.4 & 0.44 \\
Serpens SMM1 		& 25.5 & 7.1 & 0.21 \\ \hline
 & \multicolumn{3}{c}{[CI] $^3$P$_2$--$^3$P$_1$} \\
 & $\int T_{\rm{MB}} dV$ & $T_{\rm{peak}}$ & RMS$^1$ \\
 & [K km s$^{-1}$] & [K] & [K] \\ \hline
NGC 2071  & 41.8 & 4.0 & 0.80 \\
L1641 S3 MMS 1  & - & - & 1.90\\
Vela IRS 17	 & 41.2 &7.5 & 0.33 \\
Vela IRS 19  & 15.2 & 3.3 & 0.39\\
IRAS 20050+2720  & 12.0 & 4.3 &  1.30 \\
Serpens SMM1  & 10.2 & 2.9 & 0.8\\ \hline 
\end{tabular}
\end{center}
$^1$ RMS = 1$\sigma$ for channel width of 0.5 km s$^{-1}$.\\
\label{tab:total}
\end{table}
}

\def\placeTableDecomp{
\begin{table*}[!tp]
\caption{Parameters of the CO 6--5 component fits.}
\begin{center}
\begin{tabular}{l l l l l l l l l l}
\hline \hline
Source & \multicolumn{3}{c}{Broad Component} & \multicolumn{3}{c}{Medium Component} & \multicolumn{3}{c}{Narrow Component}\\ 
& $T_{\rm{peak}}$ & FWHM   &  $\int T \rm{d}V$ & $T_{\rm{peak}}$ & FWHM   &  $\int T \rm{d}V$& $T_{\rm{peak}}$ & FWHM   &  $\int T \rm{d}V$\\ 
 & (K) & (km s$^{-1}$) & (K km s$^{-1}$)& (K) & (km s$^{-1}$) & (K km s$^{-1}$)& (K) & (km s$^{-1}$) & (K km s$^{-1}$) \\ \hline
NGC 2071 & 20.1 & 27.9 & 596.9  & 51.3 & 9.3 & 508.5  & - &  - & -  \\ 
L1641 S3 MMS 1 & 2.8  & 16.0 & 47.5 & 5.9 & 5.6 & 35.0& - &  - & -\\
Vela IRS 17 & - & - & -  & 28.3 & 8.9 & 267.8 & - &  - & -   \\
Vela IRS 19 & - &  - & -  & 5.3 & 13.1 & 73.8  & 2.5 & 3.2 & 8.5 \\
IRAS 20050+2720 &  - &  -& -&12.3 & 12.1 & 159.1& - &  - & -  \\
Serpens SMM1 & -  & -  & - & 9.3 & 13.0 & 129.1 & 12.7 & 4.6 & 62.2  \\ \hline
\end{tabular}\\
\end{center}
\label{tab:decomp}
\end{table*}
}
\def\placeTableDecompCI{
\begin{table}[!tp]
\caption{Parameters of the [CI] 2--1 and $^{13}$CO 6--5 component fits.}
\begin{center}
\begin{tabular}{l l l l l l l}
\hline \hline
Source & \multicolumn{3}{c}{$^{13}$CO} \\
      & $T_{\rm{peak}}$ & FWHM  &  $\int T dV$ \\
       & (K) & (km s$^{-1}$) & (K km s$^{-1}$) \\  \hline
NGC 2071 & 13.9 & 8.0 & 118.2 \\
				  & 9.1 & 2.1 & 20.2 \\
L1641 S3 MMS 1 & 5.3 & 1.8 & 10.3 \\
Vela IRS 17 & 15.1 & 3.9 & 62.8\\
Vela IRS 19 & 0.8 & 8.7 & 7.5 \\
						 & 3.2 & 2.2 & 7.4 \\
IRAS 20050+2720 & 3.4 & 2.9 & 10.6\\
Serpens SMM1 & 0.9 & 8.0 & 8.0 \\ 
						& 6.4 & 2.6 & 17.3 \\ \hline
Source & \multicolumn{3}{c}{[CI] 2--1} \\
      & $T_{\rm{peak}}$ & FWHM  &  $\int T dV$ \\
       & (K) & (km s$^{-1}$) & (K km s$^{-1}$) \\  \hline
NGC 2071 & 3.5 & 10.9 & 40.7\\
L1641 S3 MMS 1 & \multicolumn{3}{c}{no detection} \\
Vela IRS 17 & 3.7 & 7.6 & 29.8 \\ 
            & 5.0 & 2.6 & 13.5 \\
Vela IRS 19 & 0.8 & 6.5 & 5.6 \\
 						 & 3.4 & 3.6 & 13.3 \\
IRAS 20050+2720 & 4.4 & 2.7 & 12.7 \\
Serpens SMM 1 & 2.9 & 3.2 & 9.9 \\ \hline
\end{tabular}\\
\end{center}
\label{tab:decompCI}
\end{table}
}

\def\placeTableEHV{
\begin{table}[!tp]
\caption{Velocity offsets of detected EHV components.}
\begin{center}
\begin{tabular}{l l l l l l l}
\hline \hline
Source & Velocity$^1$ & Ref $^2$ \\ 
 & (km s$^{-1}$) & \\ \hline
IRAS 20050+2720 & $\pm$40 & 1 (EHV2) \\
& N.D. ($\pm$70) & 1 (EHV1) \\
Serpens SMM1 & -28 & \\ 
& 32 & \\  \hline
\end{tabular} \\
\end{center}
\tiny 
N.D. = Not detected. Velocity in parenthesis are the velocities for known EHV emission at low-$J$ CO. \\
$^1$ w.r.t. source velocity. \\
$^2$ 1: \cite{1995ApJ...445L..51B}.
\label{tab:EHV}
\end{table}
}

\def\placeTabletau{
\begin{table}[!tp]
\caption{Derived $^{12}$CO 6--5 average optical depths in the center and line wings.}
\begin{center}
\begin{tabular}{l l l l l l l}
\hline \hline
Source & $\tau_{\rm{center}}$ & $\tau_{\rm{blue}}$ & $\tau_{\rm{red}}$ \\ \hline
NGC 2071 				 & $>$15 	& 2 & $<$ 1 \\ 
L1641 S3 MMS 1  & $>$20 		& 2	&		2			\\ 
Vela IRS 17			 & $>$15 	&	4.5 & 2.5 \\
Vela IRS 19  	 & $>$20 & - & 6.5 \\
IRAS 20050+2720 & $>$10.0 & $<$1.0 & $<$1.0 \\
Serpens SMM1 	 & $>$9.5 & 2.2 & 1.6 \\
\hline
\end{tabular} \\
\end{center}
\label{tab:tau}
\end{table}
}

\def\placeTableTemp{
\begin{table*}[!tp]
\caption{Line wing ratios of $^{12}$CO 6--5/$J_2$-$J_1$ and temperature and density estimates using RADEX. 
The density estimate is the lower 
limit for 
thermalized emission. The temperature assumes a density $n$ of 10$^5$ cm$^{-3}$.}
\begin{center}
\begin{tabular}{l l c c c c c c c}
\hline \hline
Source Name & $J_2$-$J_1$ & Ref.$^1$ &  Ratio & \multicolumn{2}{c}{Blue} & Ratio & \multicolumn{2}{c}{Red} \\ 
& & & & $T$  & $n(H_2)$ &  & $T$ & $n(H_2)$\\ 
 & & & & (K) & (10$^3$ cm$^{-3}$)  & &  (K) & (10$^3$ cm$^{-3}$)  \\ \hline
NGC 2071 & 3--2 & 1,2 & 1.0 & $>$ 100 & 300 & 1.2 & $>$100 & 100  \\
L1641 S3 MMS 1 & 3--2 & 2 & 2 & $>$70 & 500 & 2.0 & $>$70 & 400 \\
Vela IRS 17 & 1--0 & 3 & 1.5 &  90 & 9 & 1 & $>$100 & 9  \\
& 1--0 & 4 & $<$1 & 100 & 10 & $<$0.8 & $>$140 & 25 \\
Vela IRS 19 & 1--0 & 3 & $<$1 & $>$80 & 10 & 1.4 & 75 & 10   \\
& 1--0 & 4 & $<$1 & $>$80 & 10 & 1.4 & 75 & 10   \\
IRAS 20050+2720 & 2--1$^2$ & 5 & 1.5 & $>$70 &  400 &  3 & $>$50 & 200 \\
 & 2--1$^{2,3}$ & 5 & $>$10 &  $>$10 & 10 & $>$4 &  $>$50 & 75 \\
Serpens SMM1 & 4--3 & 6 & 1.7 & $>$80 & 500  & 3 & $>$60 & 100  \\
& 3--2 & 2,7,8  & 1.0 & $>$90 & 300 &1.0 & $>$ 50 & $<$300 \\
& 3--2$^3$&  2,7,8 & 2.5 & $>$60 & 100 & - & - & - \\
& 2--1 & 9,10 & 0.9 & $>$80 & 200 & 1.5 & $>$70 & $<$200 \\
\hline
\end{tabular} \\
\end{center}
\tiny $^1$ 
1: \cite{2010MNRAS.401..204B}, 2: \cite{2013A&A...553A.125S}, 3: \cite{1999A&AS..140..177W}, 
4: \cite{2007ApJ...655..316E}, 
5: \cite{1995ApJ...445L..51B}, 6: \cite{1999ApJ...513..350H}, 7: \cite{2010A&A...523A..29D}, 
8: \cite{2010MNRAS.409.1412G}, 9: \cite{1995A&A...298..594W},
10: \cite{1999MNRAS.309..141D}.\\
$^2$ at off-positions of \cite{1995ApJ...445L..51B}. \\
$^3$ EHV component at identified positions.\\
$^4$ $T_{\rm{kin}}$, assuming $n_{\rm{H}_2}$ $<$10$^5$ cm$^{-3}$. \\
\label{tab:temp}
\end{table*}
}

\def\placecorrFCO{
\begin{table}[!tp]
\caption{Correction factors used to multiply the uncorrected value with,  from \cite{Downes:2007dr}$^1$ 
for an inclination $i$ of 30 degrees. See text.}
\begin{center}
\begin{tabular}{l l l l l l l}
\hline \hline
 $t_{\rm{d}}$ & $F_{\rm{CO}}$ & 3$\sigma$ \\ \hline
0.29 & 2.8 & 1.4 \\
\hline
\end{tabular} \\
\end{center}
\label{tab:fcocorr}
\tiny $^1$ $i$ =  90-$\alpha$, as used in \cite{Downes:2007dr}.
\end{table}
}

\def\placeTablegamma{
\begin{table}[!tp]
\caption{Resulting $\gamma$ values to line luminosity.}
\begin{center}
\begin{tabular}{l l l l l l l}
\hline \hline
 & $L_{\rm{bol}}$ & M$_{\rm{env}}$ & F$_{\rm{CO}}$ \\\hline
L$_{\rm{CO}}$ & 0.91 \\
L$_{\rm{CI}}$ & 0.88\\
L$_{\rm{^{13}CO}}$ & 0.89 \\
\hline
\end{tabular} \\
\end{center}
\label{tab:gamma}
\end{table}
}

\def\placeTableFlow{
\begin{table*}[!tp]
\caption{Outflow Parameters. The dynamical time, $t_d$, and outflow force, $F_{CO}$ were corrected for inclination using the factors listed in Table 9 as well as a factor
 of 1.4 to compensate the method in determining $V_{\rm{max}}$. See text. }
\begin{center}
\begin{tabular}{l l r c c c c c c}
\hline \hline
Source & Mass & $V_{\rm{max}}$ & $R$ & $t_d$ & $\dot{M}$ &$F_{CO}$ & $L_{\rm{kin}}$ \\
& (M$_\odot$) & (km s$^{-1}$) & (10$^4$ AU) & (10$^3$ yr) & (10$^{-5}$ M$_\odot$ & (10$^{-4}$ M$_\odot$ &
 ( L$_\odot$) \\ 
 & & &  & & yr$^{-1}$) & yr$^{-1}$ km s$^{-1})$ & \\ \hline
\multicolumn{8}{c}{Red Lobes} \\ \hline
NGC 2071                 &  0.39  & 39.0  & 4.2   & 1.1  & 36.3& 145 	&  45.8   \\
L1641 S3 MMS 1           &  0.022 & 12.0  & 0.93		& 0.77 & 2.8 & 3.5 	&  0.34     \\
Vela IRS 17              &  0.27  & 10.0  & 3.5   & 3.5  & 7.7 & 7.9 	&  0.64    \\
Vela IRS 19      		   	 &  0.38  & 13.0  & 2.1   & 1.6  & 23.8 & 31.9&  3.4     \\
IRAS 20050+2720$^1$ 			 &  0.31  & 14.0  & 4.9   & 3.5  & 8.9 & 12.8 &  1.5        \\
Serpens SMM1             &  0.006	 & 12.5  & 1.3   & 1.1  & 0.5 & 0.6 &   0.06       \\
\hline
\multicolumn{8}{c}{Blue Lobes} \\ \hline
NGC 2071             	   & 0.41  & 36.5 & 42 & 1.2  & 35.3& 130  & 38.4      \\
L1641 S3 MMS 1        		 & 0.017 & 14.5 & 27 & 1.9  & 0.9 & 1.4  & 0.16           \\
Vela IRS 17              & 0.31  & 10.0 & 42 & 4.2  & 4.6 & 4.6  & 0.37         \\
Vela IRS 19$^2$          & 0.05 	& $>$3.5 & 27 & $<$7.7   & 0.7   & $>$0.23 & $>$0.007         \\
IRAS 20050+2720$^1$			 & 0.22 	& 14.0 & 42 & 3.0  & 7.4 & 10.6 & 1.2      \\
Serpens SMM1            	 & 0.005 & 16.5 & 9  & 0.5  & 1.5 & 2.6  &  0.35\\
\hline
\hline
\end{tabular} \\
\end{center}
\label{tab:flow}
\tiny
$^1$ average over the detected flows in \cite{1995ApJ...445L..51B} and B08. \\
$^2$ strongly limited by detection of V$_{\rm{max}}$.
\end{table*}
}

\def\placeFeedback{
\begin{table*}[!tp]
\caption{Comparison between three types of feedback defined in the text. }
\begin{center}
\begin{tabular}{l c c c c c}
\hline \hline
Source &  Mechanical$^1$ & \multicolumn{2}{c}{UV heating + Cluster RF $^{1,2}$} & Ratio$^3$ \\ 
& $L_{\rm{kin}}$ & $L_{\rm{CO_{narrow}}}$ & $L_{\rm{[CI]}}$ \\
&  L$_\odot$ & \multicolumn{2}{c}{[K km s$^{-1}$ pc$^2$]}  \\ \hline
NGC 2071 					& 41.2 & 3.4 & 0.78 & 9.9 \\
L1641 S3 MMS 1  	& 0.25 & 0.56 & - &  0.44 \\
Vela IRS 17 			& 0.5 & 7.9  & 4.6 & 0.04\\
Vela IRS 19 			& $<$3.4 & 2.5 & 1.6 & 0.8 \\
IRAS 20050+2720  & 1.4 & 42.8 & 2.0 & 0.03\\
Serpens SMM1 		& 0.2 & 0.3 & 0.1 & 0.5\\
\hline
\end{tabular} \\
\end{center}
\tiny $^1$ Sum of both lobes.\\
$^2$ Sum of outflow photon heating and cluster RF (radiation field), expressed in CO and [CI]. 
The CO and [CI] can contribute to both origins, see text. Values are summed over area. \\
$^3$ Ratio is defined as Entrainment/(CO Luminosity + [CI] Luminosity).
\label{tab:feedback}
\end{table*}
}


\section{Introduction}

In the current view of star formation, low-mass
($M_{\rm{star}}$ $<$3 M$_\odot$) and high-mass ($M_{\rm{star}}$ >8 M$_\odot$) star formation are described by different theories. 
Low-mass theories build on the assumption of singular collapse 
\citep{1999osps.conf..193S}, 
while various high-mass theories focus on clustering and energetic environments
\citep{2009Sci...323..754K,2014prpl.conf..149T}. The poorly studied protostars/protostellar clusters 
of intermediate mass 
($M_{\rm{star}}$ between 3 and 8 M$_\odot$, $L_{\rm{bol}}$ between 30 and 5,000 L$_\odot$) provide 
an ideal testbed for a unified theory of star formation.
Such a theory must be able to correctly address the observed level of fragmentation within 
intermediate mass protostellar regions, 
its influence on the observed properties 
and simultaneously reproduce predicted populations set by the initial mass function 
\citep{2009MNRAS.392.1363B,2010ApJ...725.1485O,Hansen:2012ib}. 

Multiplicity and fragmentation of intermediate mass protostars is a relatively unexplored area 
\citep[for the most recent reviews, see][]{2007prpl.conf..133G,Beltran:2015kn}. 
Of those studied in detail, very few deeply embedded intermediate mass protostars are truly isolated single protostars; most 
are tightly packed clusters of low-mass protostars \citep[e.g.,][]{2012ApJ...751..137V} with at times an actual intermediate mass protostar found near the center \citep[e.g.,][]{2001A&A...366..873F}. Isolated intermediate mass protostars are known, but likely very rare \citep[L1641 S3 MMS1 is the best candidate, see][]{2012ApJ...751..137V}.

Interferometric observations at submillimeter wavelengths are needed to characterize protostellar content of intermediate mass protostars, 
down to the very young, heavily embedded protostars. 
Observations with the required spatial resolution have been obtained for just a handful of cases
\citep[e.g.,][]{2001A&A...366..873F,2005A&A...444..481F,2007ApJ...667L.179T,2012A&A...540A..75F,2012ApJ...751..137V,2012ApJ...746...71C}.
Most studies lack sensitivity and/or spatial resolution to separate emission from individual protostars. 

A powerful method to directly probe star formation without time-intensive interferometric observations 
is through the study of bipolar jets and molecular outflows.
The bipolar jet not only allows part of the angular momentum to disperse and 
gravitational collapse to continue \citep{2007prpl.conf..245A}, 
it also entrains significant amounts of the surrounding gas, thereby creating 
the molecular outflow, which is thought to be the main 
source of mechanical feedback onto the parental cloud \citep{2007ApJ...662..395N}. 
Outflows also facilitate radiative feedback at larger radii, as radiation is able to escape more readily through
the lower density outflow cavities than through the protostellar envelopes.
These feedback effects dramatically affect
fragmentation and mass accretion rates \citep{2010ApJ...725.1485O,Hansen:2012ib}. 

Outflows strengths are quantified by the outflow force, $F_{\rm{CO}}$. 
Deriving outflow forces is difficult, requiring sensitive observations of the line 
wings of molecular tracers across multiple transitions.
In practice, CO is used almost exclusively.
Recent benchmarking limits variations between methods to less than an order of magnitude
\citep{2013A&A...556A..76V}. 

Outflows emerging from intermediate mass protostars have been sparsely studied. The most complete study, 
\citet[][referred to as B08 from here on]{2008A&A...481...93B} studied outflow forces of a sample of Intermediate mass outflows using a range of methods and data in comparison with detailed observations of one of them, and proposed a (cor)relation 
between the $F_{\rm{CO}}$ and total $L_{\rm{bol}}$ (see their section 6). 
This is a change from the relation between $F_{\rm{CO}}$  and individual 
luminosity, first identified by \citet{1996A&A...311..858B}.
Higher mass accretion rates on the driving source, set or influenced by the level of fragmentation,
was put forward as the origin of this effect. However, observational constraints and large uncertainties 
in the sample and the range of methods used to derive outflow forces limited validation of this hyopthesis.

Since the publication of B08, the Atacama Pathfinder 
Experiment (APEX) and Herschel Space Observatory have enabled regular
observations of spectrally resolved mid- and high-$J$ CO emission lines\footnote{Throughout this 
paper low-$J$ CO transitions are defined as having $J_{\rm{up}}$<4, mid-$J$ CO transition having $4 \leq J_{\rm{up}}\leq$ 9 and high-$J$ CO transitions having
 $J_{\rm{up}}>$9 } of
molecular outflows \citep[See][]{2009A&A...507.1425V,2009A&A...501..633V,2010A&A...518L..86F,
2010A&A...518L.121V,2010A&A...521L..40Y,2012A&A...542A..86Y,2013A&A...549L...6K,2013A&A...553A.125S}.
The reliability of
outflow force calculations has improved significantly by making use of both the 
larger range of observable CO transitions, the increase in excitation temperatures of these transitions, and
the ability to map outflows completely within a reasonable time.
One of the main results is the constraint on entrained gas temperatures of individual flows to 
50 K or higher \citep{2009A&A...501..633V,2009A&A...507.1425V}, 
warmer than the classically adopted temperatures of $\approx$30 K 
\citep[e.g.,][and many others]{2001A&A...372..899B}, but in line with shock model predictions \citep{1999A&A...344..687H}.\\

In this paper, we present new observations using the CHAMP$^+$ 
instrument mounted on APEX\footnote{This publication is based on data 
acquired with the Atacama Pathfinder Experiment (APEX). 
APEX is a collaboration between the Max-Planck-Institut fur Radioastronomie, 
the European Southern Observatory, and the Onsala Space Observatory.}
of CO and $^{13}$CO $J$=6--5 emission of outflows emerging from six intermediate mass 
protostars, 
four of which are known to form clusters of low-mass sources. 
[CI] 2--1 emission is discussed as a complement.
The goal of this paper is to validate the relation between outflow force and total luminosity 
proposed by B08 by making use of the advances of mid-$J$ CO observations and the methodology applied to low-mass protostars 
developed by this group 
\citep{2009A&A...501..633V,2009A&A...507.1425V,2012A&A...542A..86Y}.

Section 2 describes the observations, source sample and data reduction strategy.
Results are given in Section 3. Analysis is presented in section 4, while we
discuss the importance of the derived physical parameters in Section 5. Conclusions and future work
are listed in Section 6.

\section{Observations}
The dual-frequency CHAMP$^+$ array receiver \citep{2008SPIE.7020E..25G} mounted on APEX, 
was used to map the CO $J$=6--5 and $^{13}$CO $J$=6--5 transitions in six intermediate mass protostars. 
As a complement, observations of the [CI] $^3$P$_2$--$^3$P$_1$ were obtained.
Observations were carried out between November 2009 and July 2012 using the 
On-the-fly (OTF) mode. 
Fast Fourier Transform spectrometer back-ends were attached to each of 7 pixels for each frequency band,
providing a spectral resolution better than 0.1 km s$^{-1}$.
Typical system temperatures ranged between 1300 and 2000 K for the 690 GHz receivers 
and 3500 to 5000 K for 
the 810 GHz receivers.
Maps of at least 2$\farcm$5 by 2$\farcm$5 (CO $J$=6--5) 
and 1$'$ by 1$'$ ($^{13}$CO and [CI]) were obtained. Sensitivities across the maps and sample
varied by factors of 2-4 because of the different 
atmospheric conditions in combination with elevation of the sources. 
Noise levels increase by a factor of 2 at the map edges (the outer 15 $''$).
The average beam efficiency was derived to be 0.48 for the 690 GHz array and 
0.42 for the 800 GHz array. However, there are variations from month to month\footnote{see the MPIfR website 
for more information and distribution of beam efficiency measurements:
http://www3.mpifr-bonn.mpg.de/div/submmtech/heterodyne/champplus/}. Beam
efficiencies for individual scans were taken 
as close in time as possible to the observation date.

From the system temperature and beam efficiency 
measurements the total flux 
uncertainty is assumed to be 20$\%$ for the 690 GHz band and 30$\%$ for the 800 GHz band.

\placeTableSources
\subsection{Source Selection}
The targeted sample consisted of all intermediate mass protostars observable from APEX from the WISH key program on the Herschel Space Telescope
\footnote{Water in Star-forming regions with Herschel, see http://www.strw.leidenuniv.nl/WISH for more information}
\citep{2011PASP..123..138V}: 
NGC2071 \citep{2012ApJ...746...71C,2012ApJ...751..137V}, 
L1641 S3 MMS1 \citep{2012ApJ...751..137V},
Vela IRS 17 and Vela IRS 19 \citep{2005A&A...433..941G}.
The sample was completed by Serpens SMM 1 \citep[the most massive low-mass protostar included in WISH, and known to be a single protostar. See ][]{2009ApJ...706L..22V,2012A&A...542A...8K} 
and IRAS 20050+2720
(B08). Table \ref{tab:sources} lists all relevant properties (total luminosity, distance, 
total mass, $V_{\rm{LSR}}$ and estimated number of members). 
If conflicting values were reported in existing literature, 
values presented in \citet{2011PASP..123..138V}\footnote{or B08 in the case on IRAS 20050+2720} are given.

\subsection{Data Reduction}
During the observations, the raw data-streams were immediately calibrated using the APEX on-line calibrator, 
assuming an image sideband suppression of 10 dB. $^{13}$CO and [CI] observations of L1641 S3 MMS1 had to be reprocessed with the APEX
off-line calibration software owing to inaccuracies in the on-line calibration. 
Afterwards, a full reduction was done using 
standard routines in the CLASS and GREG packages of 
GILDAS\footnote{GILDAS is a set of (sub-)millimeter radioastronomical 
applications (either single-dish or interferometer) developed at 
IRAM, see http://www.iram.fr/IRAMFR/GILDAS.}. The final data product was transformed into large 
FITS cubes in main beam temperature scale with a spectral resolution of 0.1 km~s$^{-1}$.  

In the [CI] spectrum of Vela IRS 19, an absorption feature at $\sim$0 km s$^{-1}$
is present. This is caused by large-scale cloud emission at the 
off position. A spectrum taken at the off position 
revealed that the absorption is narrow and not affecting emission at the 
velocities of Vela IRS 19 which is offset by more than 10 km~s$^{-1}$.

\placeTableTotal

\placeFigureSingleCOSixFive
\placeFigureSingle13CO
\placeFigureSingleCarbon
\placeTableDecomp
\placeTableDecompCI
\section{Results}
\subsection{Line Profile of central position}
Table \ref{tab:total} presents the integrated intensities, peak temperatures and effective noise 
levels of spectra 
extracted from the central (0,0) positions, assumed to be the gravitational centers. We note that 
not all protostars 
are covered by the beam at (0,0). E.g., NGC~2071-C \citep{2012ApJ...751..137V} and 
IRAS~20050$-$2720 OVRO~2 (B08) are located over 9$''$ away from this position. Resulting spectra 
are shown in Figures \ref{fig:CO65central} (CO 6--5), 
\ref{fig:13COcentral} ($^{13}$CO 6--5) and \ref{fig:carboncentral} ([CI]).

All lines are detected with S/N $>$ 10, with the exception of a non-detection of [CI] in 
L1641 S3 MMS 1. Line profiles of the $^{12}$CO 6--5 are dominated by strong line wings, 
indicative of outflow activity. 
Absorption features are seen near the source velocities in all sources but their shapes differ from source to source. 
For NGC 2071, the absorption can be directly associated 
to large-scale material. The absorption is detected at three distinct velocities coinciding with the 
velocities of the large-scale CO, $^{13}$CO and C$^{18}$O 
3--2 emission profiles 
\citep[12, 8 and 4 km s$^{-1}$, see][]{2010MNRAS.401..204B}. As such, it is safe to assume absorptions are caused by cold material
in the outer envelope and/or large-scale cloud.
The $^{13}$CO and [CI] lines are dominated by 
narrow emission components. Wider components in these lines are only seen for Vela IRS 19 ($^{13}$CO), NGC 2071 
($^{13}$CO and [CI]) and Vela IRS 17 ([CI])

To better analyze the different components, spectra are decomposed by fitting up to three Gaussians to each profile:
a `narrow' (FWHM $<$4 km~s$^{-1}$), 
a `medium' (between 4 and 15 km~s$^{-1}$) and a `broad' ($>15$
km~s$^{-1}$) component. Absorption features are corrected for.
This method is similar to the methods used by \citet{2010A&A...521L..30K}, 
\cite{2012A&A...542A...8K} and \citet{2013A&A...553A.125S} to deconstruct 
H$_2$O and high-$J$ CO line profiles. It is possible to identify different outflow components using the `medium' 
and `broad' components and separate them 
from quiescent components traced by `narrow' components. Results of the decomposition are presented in Table \ref{tab:decomp} for CO 6--5 and Table \ref{tab:decompCI} for $^{13}$CO 6--5
and [CI] 2--1. A visual example is overplotted in Fig. \ref{fig:CO65central} on the L1641 S3 MMS 1 spectrum.

For CO 6--5, `medium' components dominate. 
Only NGC~2071 and L1641~S3~MMS1 
show `broad' components, while `narrow' emission components were found for
Vela~IRS19 and Serpens~SMM1.
 
[CI] and $^{13}$CO lines decompose into `narrow' and `medium' components, although the widths of the `medium' 
components are often lower than corresponding $^{12}$CO `medium' components. Observed variations in 
line width of `narrow' components are caused by uncertainties in the fitting routine and the achieved S/N.

\placeFlowMaps
\placeCOthirteenMaps

\subsection{Maps}

Fig. \ref{fig:flowmaps} shows the CO 6--5 emission associated with the molecular outflows
in comparison the integrated emission within 2 km s$^{-1}$ of the $V_{\rm{LSR}}$. Outflow emission was measured by
integrating between velocities of $\pm$4 to $\pm$20 km s$^{-1}$ with respect to
the source velocity. 
For the very broad flow emerging from NGC 2071, these 
cuts were changed to 
$\pm$ 10 to $\pm$40 km s$^{-1}$. If known, positions of 
(sub)millimeter detected protostars are plotted with white crosses. For the Vela sources, no interferometric (sub)millimeter
observations exist to identify individual protostars.
Figure \ref{fig:13COmaps} shows the emission of $^{13}$CO (contours) and [CI] (colors). 
Map sizes in these lines are typically smaller than the $^{12}$CO 6--5.

\section{Analysis}
\subsection{Components at central position}
The decomposition of CO 6--5 in Table \ref{tab:decomp} differs from both the CO 10--9 and 3--2 decompositions 
in \cite{2013A&A...553A.125S}. `Broad' components ($>$ 20 km s$^{-1}$ in width) 
dominate the CO 10--9, `medium' components the CO 6--5, and `narrow' components the CO 3--2. `Broad' components
are detected for CO 3--2, but are relatively weaker than the `narrow' component. Similarly, the two detected `broad' components 
are weaker than their CO 10--9 counterparts. 
From a comparison of all three decompositions we conclude that the relative contribution of the `broad' component to the total 
integrated line flux increases as a function of excitation energy.

For sources in our sample where `broad' components are detected in CO 10--9, counterparts in CO 6--5 are likely 
hidden by the noise; The focus for the CO 6--5 observations above was the size of the maps and not the sensitivity. 

$^{13}$CO emission is dominated by `narrow' emission. As seen in the contours of Fig. \ref{fig:13COmaps}, 
it originates in circumstellar envelope in most cases. The exception appears to be IRAS 20050+2720, where
$^{13}$CO peaks 30$''$ south-east of the protostars. For three sources, NGC 2071, Vela IRS 19 and Serpens SMM 1, a 
`medium' component is detected, but apart from NGC 2071, this component is much weaker than the `narrow' component. 

The [CI] emission is dominated by `narrow' emission components. Only NGC 2071 shows emission with a FWHM $>$8 km s$^{-1}$.

\subsection{Line Luminosities}
\placeLineLum
CO line luminosities show a relatively tight relation 
across the large range of luminosity and mass involved in star formation 
\citep{2005ApJ...635L.173W,2010ApJS..188..313W,2013A&A...553A.125S}. 
The line luminosity relation is defined as
\begin{equation}
~~~~~~~~~~~~~ L = 10^{\beta} L_{\rm{bol}}^{\gamma}
\end{equation}
with $L$ the line luminosity ($L_{\rm{CO}}$, $L_{\rm{[CI]}}$ or $L_{^{13}\rm{CO}}$) 
for CO 6--5, [CI] $^3$P$_2$-$^3$P$_1$ and $^{13}$CO 6--5 respectively.

Figure \ref{fig:Linelum} shows the line luminosities
compared to $L_{\rm{bol}}$, including results of CO 6--5 of low-mass protostars from \citet{2009A&A...501..633V} and \citet{2009A&A...507.1425V}. 
An average value $\gamma$ of 0.83 $\pm$0.05 is found, almost 
identical to 0.84 $\pm$0.06 of \cite{2013A&A...553A.125S} using Herschel CO 10--9 line observations 
and spanning the full range in luminosity between 1 and 10$^5$ $L_{\rm{bol}}$.
We note that even with the small sample size, [CI] line luminosities follow the expected 
correlation between low- and high-mass 
star formation through total luminosity with a slope of 0.96.

\placeTabletau
\placeTableTemp
\placeradex

\subsection{Optical depths}
Table \ref{tab:tau} presents three different optical depths derived using the 
CO 6--5/$^{13}$CO 6--5 line ratios: 
at line center ($\tau_{\rm{center}}$) and in each of the wings ($\tau_{\rm{blue}}$/$\tau_{\rm{red}}$). 
A standard $^{12}$CO/$^{13}$CO isotopologue ratio of 65 was used \citep{1994ARA&A..32..191W}.

Optical depths at line center range from 10 to $>$25, and are set by foreground absorption.
In the wings, optical depths range between $<$1 and 6.5 with most values between 1 and 3. 
We note that these are upper limits as a result of the lack of signal at higher velocities in $^{13}$CO 6--5. 
These optical depths are higher than upper limits for 
low-mass sources
($\approx$1, e.g., NGC 1333: \citealp{2012A&A...542A..86Y}, HH46: \citealp{2009A&A...501..633V}). 
Exceptions are IRAS 20050+2720 and NGC 2071, which show optically thin emission 
in the line wings.

Optical depths at different positions are consistently equal or lower than the optical depths given in Table \ref{tab:tau}.
For convenience, the optical depth for the flows are used for the full maps.

\subsection{Outflow properties}
Outflow parameters such as kinetic temperatures, densities, outflow forces and kinetic luminosities can be derived
by calculating the non-LTE radiative transfer parameters and comparing those with observed line emission.
For this sample, the off-line version of the RADEX code \citep{2007A&A...468..627V} was used provide constraints using 
ratios of the observed CO 6--5 emission over previously observed transitions 
of lower excitation (see Table \ref{tab:temp}). Diagnostical plots for ratios with respect to CO 6--5 are presented in Fig. \ref{fig:radex}.

\subsubsection{Temperature and density}
Diagnostical plots produced by RADEX as shown in \cite{2007A&A...468..627V} reveal that ratios provide solutions with
a degeneracy between temperature and density.
To break this degeneracy other molecular tracers are required to independently derive excitation constraints.
CO line ratios covering three or more transitions can provide additional information, 
but are often insufficient to completely solve the degeneracy.
However, ratios using the CO 6--5 line emission are able to exclude large areas of the parameter space. 
E.g., temperatures under 50 K are found to be excluded for many outflows
\citep{2009A&A...501..633V,2009A&A...507.1425V,2012A&A...542A..86Y}. 
For a more thorough discussion on RADEX solutions concerning CO and including thermal versus sub-thermal 
excitation, we refer the reader to \citet{2012A&A...542A..86Y}.

To derive the excitation parameters of this sample, 
RADEX was run in the optically thin limit, adopting the following parameters: a line width of 10 km s$^{-1}$, 
a column density of 10$^{12}$ cm$^{-2}$, and a background radiation field of 2.73 K.

In turn, the excitation conditions were investigated by considering two scenarios.
First, (lower limits to) the temperatures were derived by assuming a density of 10$^5$ cm$^{-3}$.
Second, a lower limit for the density is given. This solution is the lowest density at which emission is fully thermalized. 
In other words, the density given is the lowest density for which the ratio solely depends on temperature (the limits found at the right sides of the 
diagnostical plots in Fig. \ref{fig:radex} where lines are horizontal).
These two scenarios were chosen as they likely best reflect the true physical conditions.
Results can be found in Table \ref{tab:temp}.

Temperature limits of 50 K are found for all flows\footnote{The limit of 10 K for the blue flow of IRAS 20050 was derived 
with a synthesized beam and is likely suffering from filtered out large-scale emission.}. 
These are in agreement with temperature constraints for 
low-mass protostars \citep{2009A&A...501..633V,2009A&A...507.1425V,2012A&A...542A..86Y} 
Technically, lower temperatures are not excluded, but require densities 
 $>$ 10$^6$ cm$^{-3}$. Spherical envelope modelling restricts such densities to
$<$2000 AU from the central protostar \citep[See Dusty models of ][]{2012A&A...542A...8K}. 
Similarly, B08 restricted densities in IRAS 20050+2720 to $1.3-3 \times$ 10$^6$ cm$^{-3}$ at
radii $<$3000 AU. In general, densities at larger radii are significantly lower,
with most cloud densities derived to be on the order of 10$^4$ cm$^{-3}$. Compression factors can be invoked to compensate for this change in density (i.e, local density enhancements). However, compression factors of more than three orders of magnitude
are required to keep CO emission thermalized along the entire observed flow ($>$20,000 AU). 
Molecular tracers such as H$_2$O \citep{2012A&A...538A..45S}
or HCO$^+$ \citep{2009A&A...501..633V} independently provide density constraints of $<$10$^6$ cm$^{-3}$, 
limiting compression factors to two orders of magnitude.
As such, the sub-thermal low-$n$, high-$T$ solution is the preferred solution.

\subsubsection{The optically thin limit}
Tests were carried out to verify the optically thin assumption used above and the effect higher optical depths would have on the density and temperature derivations. This was done by using
 significantly higher column densities (10$^{15}$-10$^{17}$ cm$^{-2}$) in the RADEX simulations.
These optically thick solutions provide significantly higher constraints on temperature ($>$200 K instead of $>$50 K), while the effective $n_{\rm{crit}}$ 
was found to increase, thus increasing the lower limit on the density given in Table \ref{tab:temp}. 
Since the derived optical depths for outflowing gas are mostly upper limits, adopting the optically thin RADEX solutions provides us with the most
conservative, but likely more realistic, estimate. 

\subsubsection{Velocity and spatial variations}
Line ratios were found to be relatively constant in velocity, 
with ratio variations typically on the order of 20 to 40$\%$.
This is similar to the behaviour of CO in flows around HH~46 and NGC 1333 IRAS 2 
\citep[Fig. 10 in][]{2009A&A...501..633V,2012A&A...542A..86Y}.

Line ratios at other positions also show little to no significant difference.
It can thus be concluded that variations in the excitation mechanisms along the large-scale flows are small. 
\placeTableFlow

\placecorrFCO
\subsubsection{Mass outflow rate and outflow force}
Outflow forces and kinetic luminosities are derived using 
the H$_2$ column density. 
Column densities are derived using Eq. 1 of \cite{1998ApJ...502..315H} using parameters for the CO 6--5 transition:
\begin{equation}
N = 10^5\frac{3\kappa^2}{4h \pi^3\nu^2\mu^2}(e^{\frac{h\nu J_l}{2 \kappa T} })\frac{T + 
\frac{h \nu}{6\kappa(J_l + 1)}}{e^{-h\nu /\kappa T}}\int T_{mb} \frac{\tau}{1-e^{-\tau}} dV
\end{equation}
where $\kappa$ is the Boltzmann constant, $h$ the Planck constant, $\mu$ the permanent 
dipole moment (0.122 Debye for CO), $\nu$ the 
frequency of the transition, $J_l$ the quantum number of the lower rotational state, and 
$ \int T_{mb} (\tau/(1-e^{-\tau})) dV $ 
the integrated line intensity, corrected for optical depth. All quantities are in cgs units, except for the velocity which is in km s$^{-1}$. 
The total mass is calculated by summing column densities across the map, assuming a H$_2$/CO ratio of 10$^4$.

The outflow force, $F_{CO}$, is derived using the integrated intensity as a function of velocity, corrected for optical depth and subsequently integrated over the observed area of pixels $i$ and in turn corrected for the inclination. From recent benchmarking, the most reliable method to calculate outflow forces is the 'M7' or 'separation' method \citep{2013A&A...556A..76V}. 
In this method the dynamical age and force of a flow are considered to be independent quantities. 
The dynamical age, $t_{\rm{d}}$ is defined as the measured radius, $R$, divided by $V_{\rm{max}}$. 
Using the intensity weighted velocities the outflow force is thus expressed with the following equation:
\begin{equation}
F_{CO} = c \times \frac{K (\sum_{i} [ \int T_{mb}  \frac{\tau}{1-e^{-\tau}} V'dV']_i) V_{\rm{max}}}{R_{\rm{lobe}}}
\end{equation}
Here $c$ is the inclination correction and $K$ the temperature-dependent correction factor. $R_{\rm{lobe}}$ is the radius of the lobe. For more information on this method, see \citet{2013A&A...556A..76V}.
The correction factor is derived from the values of Table 6 of \cite{Downes:2007dr} (see Table 7).

Owing to atmospheric effects, observations at 691 GHz cannot be obtained with a similar signal to noise ratio as 
its low-$J$ counterparts within reasonable times. As such, 
the M7 method was changed on three points w.r.t. \cite{2013A&A...556A..76V}:

\begin{enumerate}
\item Densities derived from line ratios were found to be 10$^4$ cm$^{-3}$ or 
higher. Therefore, the $\eta$ = 1 case of \cite{Downes:2007dr} is used instead of  
the geometric mean of 0.1 and 1 adopted by \cite{2013A&A...556A..76V}.
The latter corresponds to densities below 10$^4$ cm$^{-3}$.
As before, these low densities would imply unlikely temperatures of 200 K or higher.
Although not excluded, typical envelope models \citep{2012A&A...542A...8K} 
already have higher densities. 
In addition, extrapolation of the correction factors from \cite{Downes:2007dr} above 100 K is not reliable. 
\item $\Delta V_{\rm{max}}$ was measured using a 3$\sigma$ limit, 
instead of a 1$\sigma$ limit. Data quality at 690 GHz was found to be insufficient to make a reliable 1$\sigma$ 
limit derivation.  
With the higher system temperatures of the CHAMP$^+$ due to the lower atmospheric transmission at 690 GHz, the effective S/N of the CO 6--5 observations here
are a factor of 5 or more 
lower than for CO 3--2 used by \cite{2013A&A...556A..76V}.
To avoid any potential systematic errors introduced by the data quality 
but still correctly approach true values for $\Delta V_{\rm{max}}$ correctly, both 
$t_{\rm{d}}$ and $F_{\rm{CO}}$ were corrected with an additional factor of 1.4. 
This was tested by extrapolating the gaussian fits to $V_{\rm{max}}$ 
and found to be robust. The correction factor was derived from tests using the appendix
of \cite{2013A&A...556A..76V} as well as similar tests on data presented here and \cite{2009A&A...501..633V}.
$t_d$ is divided by this correction factor, while $F_{\rm{CO}}$ is multiplied. 
\item No reliable information on individual viewing angles of the outflows
with respect to the plane of the sky is 
available. An average value of 32 degrees for the angle of the outflow with the plane of the sky is adopted. This is the expected mean value for a randomly distributed sample of outflow inclinations.
\end{enumerate}

Table \ref{tab:fcocorr} lists the final correction factors used.
As a complement to the outflow force, the kinetic luminosity of the flows, $L_{\rm{kin}}$, was calculated using 
\begin{equation}
~~~~~~L_{\rm{kin}} = F_{\rm{CO}} \times V_{\rm{max}}/2. 
\end{equation} 

Table \ref{tab:flow} lists 
the final values for all outflow parameters. 
$V_{\rm{max}}$ values of both lobes are consistently of similar value, with the exception of the blue side of 
Vela IRS 19, which is not detected strongly. Its derived parameters are considered
lower limits. 
Dynamical times are a factor of 2 shorter than the average for the low-mass protostar sample \citep{2009A&A...508..259V,2013A&A...556A..76V}. 
This may be indicative that no evolved intermediate mass protostellar cluster was included.
Outflow forces are factors of 10 to 300 higher than low-mass protostars
\citep{1996A&A...311..858B}. When compared to the results of \cite{2013A&A...558A.125D}, the outflow forces found are consistent with the low-end of values for high-mass sources (10$^{-4}$ M$_{\odot}$ yr$^{-1}$ km s$^{-1}$ or higher for luminosities of 100 and higher). 

\placeFigureFcoLum
\placeFigureFcoMenv
\subsubsection{The multiple flows of IRAS 20050+2720}
B08 observed IRAS 20050+2720 using a resolution of 3 arcseconds. 
This allows one to identify all individual outflows. It is seen that the observations of interferometers do not resolve out any emission. The total outflowing mass and outflow force co-added for both 
flows in IRAS 20050+2720 add up to 
0.25 M$_\odot$ and 6.4 10$^{-4}$ M$_\odot$ km s$^{-1}$ yr$^{-1}$ (B08). 
Co-added outflow forces from CO 6--5 are less than a factor of 3 higher. With the systematic difference due to the difference in temperature (29 K for B08, 50 K for this study), this factor is well within the assumed 
inaccuracy of the M7 method \citep{2013A&A...556A..76V}. 

\section{Discussion}
\subsection{Does fragmentation enhance outflow forces? }

Using a large sample of outflows emerging from low-mass protostellar environments, \citet[][from here on referred to as B96]{1996A&A...311..858B} revealed a relation between the bolometric 
luminosity of the driving source and the force of its outflow (from here on referred to as the B96 relation). 
The relation inherently has a relatively large scatter, and the influence of evolutionary effects could not be determined accurately. Since then, 
many other studies corroborated this relation, and reveals it may apply across the many orders of magnitude in luminosity in star formation, up to and including high-mass star formation \citep{2013A&A...558A.125D}.
Figure \ref{fig:fcolum} shows the B96 relation (red line) in comparison with the results obtained here (black filled points) and B08 (white diamonds). Studies of low-mass 
protostars using CO 6--5 and/or similar methodology to derive outflow forces are included as reference \citep[][grey symbols]{2009A&A...507.1425V,2013A&A...556A..76V,2015A&A...576A.109Y}. For the intermediate mass flows, the outflow force at first sight correlates with the total $L_{\rm{bol}}$ with the B96 relation, similar to the conclusions of B08.
The largest deviation from the relation  is the NGC 2071 flow, where the outflow force is an order of magnitude higher than expected based on the B96 relation but still well within the observed scatter.

Interestingly enough, the median of the the outflow forces in \cite{2015A&A...576A.109Y} as derived for low-mass sources using CO 6--5 is also statistically significant above the B96 relation. 
An excess of an order of magnitude for fragmented intermediate mass sources was identified earlier by B08, although their results were inferred from data with a  
 higher uncertainty on the outflow forces.
Direct calculations of gas temperatures, correct derivations of optical depths and 
a uniform derivation of $F_{\rm{CO}}$ have improved the reliability of the derived values. 
The observed excess of B08 can thus neither be corroborated or invalidated. 
What is more likely is that with the improvements made in observations, the better understanding of the derivation of outflow forces and the access to tracer lines better suited to track entrained outflow material (in this case the CO 6--5), outflow forces are slightly higher than the original B96 relation, although the most important aspect of it, the slope, remains the same.

Figure \ref{fig:fcomenv} compares the $F_{\rm{CO}}$ and $M_{\rm{env}}$ relation of
\cite{1996A&A...311..858B} for the same set of samples above with the B96 relation shown as a red line. The correlation between $M_{\rm{env}}$ and $F_{\rm{CO}}$ is clearly visible for intermediate mass sources and in direct agreement with those derived from low-mass sources.
It should be noted that compared to the low-mass sources (shown in white, light gray and dark gray), the scatter for the
intermediate mass sources (shown in black or white diamonds) has increased by a factor 2. 
Most likely this enhanced scatter is caused by measurement uncertainties due to intermediate mass sources being more distant. Intermediate mass sources are a factor of 2 to 5 more distant than
typical low-mass sources. From these results, the results of B08 cannot be corroborated. The observed enhancement in outflow forces seen by B08 are reproduced, but are within the scatter of the B96 relation. In addition, the advancement of more accurate observations and differences between them \cite{2013A&A...556A..76V}and corresponding constraints can be invoked to explain any changes .

\subsection{Neutral carbon}
Neutral carbon has long been assumed  to be created from 
interactions of the gas with the ISRF (Interstellar Radiation Field), which produces atomic gas components in a PDR scenario 
from photo-dissociation of carbon-bearing 
species \citep{1997ARA&A..35..179H}, although atomic gas within the outflow cavity surface PDRs could contribute 
to neutral carbon emission \citep{1997ARA&A..35..179H}. 

Indeed, the observed spatial distribution of the [CI] $^3$P$_2$-$^3$P$_1$ emission is clearly different from that of CO. In addition, no correlation
to the direction and/or strength of the outflow is detected.  
Figure \ref{fig:13COmaps} revealed [CI] emission to be smoothly distributed over the circumcluster envelope, with concentrations at or near  
the protostellar positions.
The only correlation seen for [CI] is in the line luminosity, which reproduces the same slope as derived for CO and $^{13}$CO 6--5. This is not surprising as [CI] is clearly expected to be coupled to CO, which scales linearly with luminosity.
Whether or not this correlation is evidence for atomic gas emission being correlated with outflow activity itself cannot be confirmed owing to the low number of detections of [CI].

\section{Conclusions}

This paper presents new spectral line observations of six protostellar clusters of intermediate mass, 
with total luminosities ranging from 30 L$_\odot$ to $\approx$750 L$_\odot$. 
CO $J$=6--5, $^{13}$CO $J$=6--5 and [CI] spectrally resolved maps were obtained 
with the CHAMP$^+$ instrument on APEX.
Using line decomposition, accurate optical depths and line luminosity relations, 
densities, temperatures, forces and kinetic luminosities of the molecular outflows were derived and presented. 
The conclusions can be summed up as follows:
\begin{itemize}
\item The CO 6--5 line profiles are dominated by outflow 
related emission, but show quiescent emission in the cloud as well.
\item Mid-$J$ CO line luminosities adhere to the correlation between total luminosity and 
line luminosity identified by \cite{2013A&A...553A.125S} for low- and high-$J$ CO. 
\item There is no corroboration of the result presented in  \citet{2008A&A...481...93B} that proposed an apparent enhancement in outflow force for fragmented intermediate mass sources. Although an enhancement of the outflow forces as a function of total bolometric luminosity is seen in comparison with the original B96 relation, this increase can also be attributed to methodology or the improvement in temperature and density derivations due to the inclusion of mid-$J$ CO. 
\end{itemize}
Future work on outflows emerging from protostellar clusters of intermediate mass require observations down to scales of individual protostars. 
Properties of individual sources are necessary to draw conclusions about the influence of fragmentation. Although near and mid-infrared observations 
have been acquired that can be used (e.g., Spitzer, WISE), (sub)millimeter observations with sufficient spatial and spectral resolution are rare.
ALMA is able to routinely do such observations in minutes through several CO transitions, although other interferometers at these wavelengths (SMA, CARMA, IRAM Plateau de Bure and 
its successor NOEMA) should not be discounted even though their lack of access to mid-$J$ CO lines will limit their effectiveness.  The GREAT instrument on SOFIA and in particular the upGREAT array extension of the instrument may spectrally resolve foreground atomic gas from outflowing atomic gas. 
In combination with ALMA Band 8 and 10, such observations must be used to interpret the [CI] observations.

\begin{acknowledgements}
TvK is supported by the Allegro ARC node in Leiden and NOVA (Nederlandse Onderzoeksschool voor Astronomie) and 
NWO (Nederlandse Organisatie voor Wetenschappelijk Onderzoek).
RJvW acknowledges support provided by NASA through the Einstein Postdoctoral grant number PF2-130104 awarded by the Chandra X-ray Center, which is
operated by the Smithsonian Astrophysical Observatory for NASA under contract NAS8-03060. AK acknowledges support from  
the Polish National Science Center grant 2013/11/N/ST9/00400 and the Foundation for Polish Science (FNP).
Construction of CHAMP$^+$ was a collaboration between the Max-Planck-Institut für Radioastronomie Bonn, 
Germany; SRON Netherlands Institute for Space Research, Groningen, the Netherlands; 
the Netherlands Research School for Astronomy (NOVA); and the Kavli Institute of 
Nanoscience at Delft University of Technology, the Netherlands; with support 
from the Netherlands Organization for Scientific Research (NWO) grant 600.063.310.10. 
We thank the APEX staff, in particular the scientists 
(Per Bergman, Andreas Lundgren, Michael Dumke, Francisco Montenegro and Rodrigo Parra) 
and operators (Francisco `Pancho' Azagra, Claudio Agurto, Felipe Mac Auliffe, 
Paulina Venegas and Mauricio Martinez) for their warm welcome over the years.  
\end{acknowledgements}
 
\bibliographystyle{aa}
\bibliography{im2}
\end{document}